\documentclass[preprint,showpacs,preprintnumbers,amsmath,amssymb,prb]{revtex4-1}

\usepackage{graphicx}% Include figure files
\usepackage{dcolumn}% Align table columns on decimal point
\usepackage{bm}% bold math
\usepackage{bigints}%[bigints.sty]
\usepackage{theorem}
\newcommand{\vecvar}[1]{\mbox{\boldmath$#1$}}
\newcommand{\e}{\mbox{e}}

\newcommand{\dsla}{\hspace{-0.5mm}/\hspace{-0.5mm}/} %double slash
\newcommand{\neareq}{~$=$\hspace{-.7em}\raisebox{1.1ex}{.}\hspace{.1em}\raisebox{-0.2ex}{.}}
\begin{document}

\preprint{PRE/003}

\title{First-principles calculation method for electron transport based on grid Lippmann--Schwinger equation}% Force line breaks with \\

\author{Yoshiyuki Egami}
\affiliation{Division of Applied Physics, Faculty of Engineering, Hokkaido University, Sapporo, Hokkaido 060-8628, Japan}

\author{Shigeru Iwase}
\affiliation{Division of Precision Science \& Technology and Applied Physics, Graduate School of Engineering, Osaka University, Suita, Osaka 565-0871, Japan}

\author{Shigeru Tsukamoto}
\affiliation{Peter Gr\"unberg Institut \& Institute for Advanced Simulation, Forschungszentrum J\"ulich and JARA, D-52425 J\"ulich, Germany}

\author{Tomoya Ono}
\affiliation{Center for Computational Sciences, University of Tsukuba, Tsukuba, Ibaraki 305-8577, Japan \\
JST-PRESTO, Kawaguchi, Saitama 332-0012, Japan}

\author{Kikuji Hirose}
\affiliation{Graduate School of Engineering, Osaka University, Suita, Osaka 565-0871, Japan}

\date{\today}% It is always \today, today,

\begin{abstract}
We develop a first-principles electron-transport simulator based on the Lippmann--Schwinger (LS) equation within the framework of the real-space finite-difference scheme. In our fully real-space based LS (grid LS) method, the ratio expression technique for the scattering wave functions and the Green's function elements of the reference system is employed to avoid numerical collapse. Furthermore, we present analytical expressions and/or prominent calculation procedures for the retarded Green's function, which are utilized in the grid LS approach.
In order to demonstrate the performance of the grid LS method, we simulate the electron-transport properties of the semiconductor/oxide interfaces sandwiched between semi-infinite metal electrodes.
The results confirm that the leakage current through the (001)Si/SiO$_2$ model becomes much larger when the dangling-bond (DB) state is induced by a defect in the oxygen layer while that through the (001)Ge/GeO$_2$ model is insensitive to the DB state.
\end{abstract}

\pacs{
}% PACS, the Physics and Astronomy
\maketitle

%!!!!!!!!!!!!!!!!!!!!!!!!!!!!!!!!!!!!!!!!!!!!!!!!!!!!!!!!!!!!!!!!!!!!
\section{\label{sec:level1}Introduction}
Electron-transport calculations are important tools to investigate and develop materials for new electronic devices. Recently, to obtain more practical knowledge on electron-transport properties of nanoscale structures, long-range and large-scale transport simulations have attracted much interest. However, such simulations are very hard task since huge computational costs growing with a system size are required. Therefore, it is important to develop an efficient electron-transport simulator.

The Lippmann--Schwinger (LS) equation method proposed by Lang {\it et al.}~\cite{LS_lang2,LS_lang1,LS_lang3} is one of popular methods, which enable us to obtain the scattering wave functions of nanoscale structures sandwiched between electrodes by solving the integral equation of the second-kind Fredholm equation. 
When the reference system consists of only bare left and right electrodes with the empty transition region, scattering wave functions can be efficiently evaluated for a variety of structures of nanoscale junctions set up in the transition region by using the same reference Green's function of the bare electrode system, where the computation of the reference Green's function has only to be performed once.
Moreover, for a similar reason, the LS equation is utilized in the implementation of self-consistent calculations for the convergence of electronic states in infinitely open systems~\cite{LS_kobayashi1,LS_tsuka1,LS_tsuka2,tsukaLS,LS_kobayashi2}.
In the conventional LS equation method, scattering wave functions are expressed in the Laue representation, that is, the LS equation is solved by using a 2-dimensional plane-wave expansion in the directions parallel to the electrode surface (lateral directions) and a real-space discretization of the coordinate in the direction perpendicular to that (longitudinal direction).
In the LS equation method, however, one may frequently encounter a numerical difficulty such that a part of the Green's function expressed in a variable-separable form drastically varies due to the appearance of evanescent waves exponentially growing and decaying in the longitudinal direction. To overcome this issue, in the previous study~\cite{tsukaLS}, we proposed the procedure of the ratio expression for the Green's function matrix elements in the Laue representation as a remedy for avoiding the numerical collapse.

So far, we developed the several simulators to elucidate the electronic properties of nanostructures based on the real-space finite-difference~(RSFD) approach~\cite{ono1,icp,fujimoto,iobm,sasaki,ono2,ono3,ono4,ono5,onoNEGF,tsukaLS}, in which the system is divided by equally spaced grid points, within the framework of the density functional theory~\cite{dft,kohn}. For electron-transport simulations, the RSFD method has several advantages compared with the method of the Laue representation from the fundamental and practical points of view. Firstly, the finite differentiation for the kinetic-energy operator is treated on the {\it equal footing} in all three directions. This avoids numerical errors due to the artificial anisotropy between the lateral and longitudinal directions at any grid spacing. Secondly, the computational accuracy can be improved by employing a higher-order finite-difference formula. Thirdly, in the lateral directions, isolated boundary conditions are available as well as periodic ones, which enable us to treat electrodes as leads. Furthermore, the algorithm of the RSFD method is suitable for massively parallel computing\cite{RSDFT}.

In this paper, we present the fully real-space based LS method and the ratio expression technique for the Green's function of the reference system within the approach of the RSFD. This method is referred to the grid LS method.
To demonstrate the performance of the grid LS method, we use it to investigate the electron-transport properties of the (001)Si/SiO$_2$ and (001)Ge/GeO$_2$ models connected to semi-infinite electrodes. We also estimate how the dangling bond~(DB) caused by an oxygen vacancy contributes to leakage currents across the interface between the semiconductor and oxide. The results indicate that the leakage current attributed to the DB state in the Si/SiO$_2$ model is much larger than that in the Ge/GeO$_2$ model.

In the followings of this paper, Section~\ref{sec:level2} gives details of the computational scheme used to develop the grid LS method. Section~\ref{sec:level3} presents a demonstration of our method, in which we use it to examine transport properties of Si/SiO$_2$ and Ge/GeO$_2$ models and to reveal how the leakage current is influenced by the DB state that arises due to an oxygen vacancy. Conclusions are given in Section~\ref{sec:level4} and mathematical details are described in Appendices~A and B.

%!!!!!!!!!!!!!!!!!!!!!!!!!!!!!!!!!!!!!!!!!!!!!!!!!!!!!!!!!!!!!!!!!!!!
\newpage
\section{\label{sec:level2}Computational Formalism}
We propose an efficient procedure to obtain the solution of Kohn--Sham equation for a system where the nanoscale junction is sandwiched between semi-infinite electrodes within the framework of the RSFD scheme. The effective potential is close to periodic bulk potentials as it goes deeply inside the left and right electrodes, so that the whole infinite system can be appropriately divided into three parts: the left electrode, the transition region, and the right electrode. The Hamiltonian of the system, $H$, is defined by
\begin{eqnarray}
H=-\frac{1}{2}\nabla^2+v(\vecvar{r},\vecvar{r}^\prime),
\label{eqn:ls001-01}
\end{eqnarray}
with
\begin{eqnarray}
v(\vecvar{r},\vecvar{r}^\prime) &=& \{v_h(\vecvar{r})+v_{xc}(\vecvar{r})+v_l(\vecvar{r})\}\delta(\vecvar{r}-\vecvar{r}^\prime)+v_{nl}(\vecvar{r},\vecvar{r}^\prime),
\label{eqn:ls001-02}
\end{eqnarray}
where $v_h(\vecvar{r})$ and $v_{xc}(\vecvar{r})$ are the Hartree and exchange-correlation potentials, respectively, and $v_l(\vecvar{r})$ and $v_{nl}(\vecvar{r},\vecvar{r}^\prime)$ are local and nonlocal parts of atomic pseudopotentials, respectively.

Assuming that the Hamiltonian in the transition region can be decomposed into an unperturbed part $H^0$ and a perturbation $\delta v(\vecvar{r},\vecvar{r}^\prime)=H-H^0$, we rewrite the Kohn--Sham equation as
\begin{eqnarray}
(E-H^0)\psi(\vecvar{r}) &=& \int_{-\infty}^{\infty}\!\!\!\!\!d\vecvar{r}^\prime \delta v(\vecvar{r},\vecvar{r}^\prime)\psi(\vecvar{r}^\prime),
\label{eqn:ls001}
\end{eqnarray}
where $\psi(\vecvar{r})$ is the scattering wave function for an incident wave coming from the left or right electrode with the energy $E$. The subscript 0 on the variables indicates that they are evaluated in the unperturbed reference system. Here, for convenience, we assumed $H^0$ not to contain the nonlocal parts of the pseudopotentials. Once the retarded Green's function $g^{r0}_T(\vecvar{r},\vecvar{r}^\prime;E)$ in the transition region associated with the unperturbed part $H^0$ is known, Eq.~(\ref{eqn:ls001}) is put into the LS equation in a form of the integral equation, i.e.,
\begin{eqnarray}
\psi(\vecvar{r}) &=& \psi^0(\vecvar{r}) + \int \hspace{-3mm} \int d\vecvar{r}^\prime d\vecvar{r}^{\prime\prime} g^{r0}_T(\vecvar{r},\vecvar{r}^\prime;E) \delta v(\vecvar{r}^\prime,\vecvar{r}^{\prime\prime}) \psi(\vecvar{r}^{\prime\prime})
\label{eqn:ls002}
\end{eqnarray}
with the unperturbed wave function $\psi^0(\vecvar{r})$.
Equation~(\ref{eqn:ls002}) provides a unified treatment of the Kohn--Sham equation and the boundary conditions\cite{LS_lang1,LS_lang2,LS_kobayashi1,LS_kobayashi2,LS_lang3,LS_tsuka1,LS_tsuka2,tsukaLS}. In the case where the incident Bloch wave $\phi^{in}(\vecvar{r}_{\dsla},z)$ propagates from deep inside the left electrode, the boundary condition is
\begin{eqnarray}
\label{eqn:ls003}
\psi(\vecvar{r}_{\dsla},z) &=& 
\left\{
 \begin{array}{ll}
  \phi^{in}(\vecvar{r}_{\dsla},z)+ {\displaystyle{\sum_{j=1}^{N}}}r_j\phi^{ref}_j(\vecvar{r}_{\dsla},z) & $ in the left electrode$ \\
  {\displaystyle{\sum_{j=1}^{N}}}t_j\phi^{tra}_j(\vecvar{r}_{\dsla},z) & $ in the right electrode$
 \end{array}
\right. .
\end{eqnarray}
Here, $\phi^{ref}_j(\vecvar{r}_{\dsla},z)$ is a reflected wave that propagates and decays into the left electrode, $\phi^{tra}_j(\vecvar{r}_{\dsla},z)$ within the right electrode is a transmitted wave, and $r_j$ and $t_j$ are unknown reflection and transmission coefficients, respectively. The lateral ($x$ and $y$) and $z$ directions are set to be parallel and perpendicular to the electrode surface, respectively. The system is assumed to be periodic in the lateral direction and infinite in the $z$ direction. The case of incident electrons coming from the right electrode can be considered in the same manner.

\begin{figure}[p]
%\begin{figure}[htb]
\begin{center}
\includegraphics*[width=16.0cm]{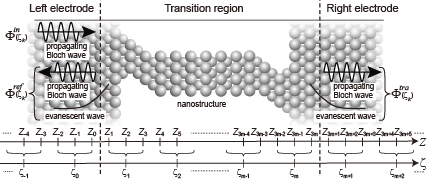}
\caption{ 
Sketch of the relationship between $z$ and $\zeta$ in the computational model in the case of $\mathcal{N}_f=3$. 
The system consists of the transition region\index{transition region} sandwiched between the left and right semi-infinite crystalline electrodes. In the left electrode, the incident wave and the reflected waves consisting of the propagating and evanescent ones are illustrated by $\Phi^{in}(\zeta_k)$ and $\Phi^{ref}(\zeta_k)$, respectively, while in the right electrode, the transmitted waves composed of the propagating and decaying evanescent ones toward the right side are denoted by $\Phi^{tra}(\zeta_k)$. Here, $x(y)$ and $z$ are coordinates perpendicular and parallel to the nanoscale junction, respectively.
}
\label{fig:zeta_coord}
\end{center}
\end{figure}

In this paper, the LS equation is solved within the RSFD scheme~\cite{icp}. The RSFD approach enables us to treat arbitrary boundary conditions and to calculate the atomic and electronic structures to high accuracy.
The whole system is composed by the transition region sandwiched between semi-infinite left and right electrodes and is divided by grid points with an equal spacing of $h_\mu=L_\mu / N_\mu$, where $L_\mu$ and $N_\mu$ are the length and the number of grid points in the $\mu$ direction ($\mu=x$, $y$, and $z$) of the transition region, respectively.
Here, we assume a 2-dimensional periodicity in the lateral directions and employ a generalized z coordinate $\zeta_k$ instead of $z_k$, which stands for the group index of $z$ coordinates within the closed interval $[z_{(k-1)\mathcal{N}_f+1},z_{k\mathcal{N}_f}]$, where $\mathcal{N}_f$ is the number of $x$--$y$ grid planes involved in $\zeta_k$ (see Fig.~\ref{fig:zeta_coord}); $\mathcal{N}_f$ corresponds to the order of the finite-difference approximation for the kinetic energy operator in the Kohn--Sham equation\cite{fd,rsfd1} and is chosen so as to include the nonlocal region of pseudopotentials to obtain highly accurate results.

The Kohn--Sham equation is written in a discretized matrix form~\cite{ono1,icp,fujimoto,iobm,sasaki,ono2,ono3,ono4,ono5,onoNEGF,tsukaLS} as
\begin{eqnarray}
-B(\zeta_{k-1};\vecvar{k}^B_{\dsla})^{\dagger} \Psi(\zeta_{k-1};\vecvar{k}^B_{\dsla}) + \left[E-H(\zeta_k;\vecvar{k}^B_{\dsla})\right]\Psi(\zeta_{k};\vecvar{k}^B_{\dsla}) -B(\zeta_{k+1};\vecvar{k}^B_{\dsla}) \Psi(\zeta_{k+1};\vecvar{k}^B_{\dsla})= 0 &, & \nonumber \\
(k=-\infty,...,-1,0,1,...,\infty) & &
\label{eqn:ls01}
\end{eqnarray}
where $H(\zeta_k;\vecvar{k}^B_{\dsla})$ and $B(\zeta_k;\vecvar{k}^B_{\dsla})$ denoting the $N$-dimensional block matrices ($N=N_x \times N_y \times \mathcal{N}_f$) are the diagonal and off-diagonal elements of the Hamiltonian block-tridiagonal matrix $\hat{H}(\vecvar{k}^B_{\dsla})$, respectively.
$H(\zeta_k;\vecvar{k}^B_{\dsla})$ includes the potential on the $x$--$y$ planes at $\zeta=\zeta_k$, $\Psi(\zeta_k;\vecvar{k}^B_{\dsla})$ is a set of the $N$ values of the wave functions on the $x$--$y$ planes at $\zeta=\zeta_k$, and $\vecvar{k}^B_{\dsla}=(k^B_x,k^B_y)$ is the lateral Bloch wave vector within the first Brillouin zone. Hereafter, for simplicity, $\vecvar{k}^B_{\dsla}$ is ignored throughout.
We assume that the Hamiltonian in the transition region can be decomposed into an unperturbed part $H^0(\zeta_k)$ and a perturbation ${\delta}V(\zeta_k,\zeta_l)$ as well as in the case of the non-discretized treatment mentioned above. When the electrodes in the unperturbed reference system are adopted to be exactly those in the perturbed system described by $H(\zeta_k)$, the perturbation ${\delta}V(\zeta_k,\zeta_l)$ has nonzero elements only in the transition region $(\zeta_1 \leq \zeta_{k(l)} \leq \zeta_{m})$ as 
\begin{eqnarray}
\label{eqn:ls02-2}
\delta V(\zeta_k,\zeta_l) &=& 
\left\{
 \begin{array}{ll}
  H(\zeta_k)-H^0(\zeta_k) &\quad (k=l) \\
  B(\zeta_k)-B^0(\zeta_k) &\quad (k=l-1) \\
  B(\zeta_{k-1})^{\dagger}-B^0(\zeta_{k-1})^{\dagger} &\quad (k=l+1) \\
  0 &\quad $otherwise$
 \end{array}
\right. .
\end{eqnarray}
Now, by using the discretized retarded Green's function ${G}^{r0}_T (\zeta_k,\zeta_l;E)$ in the transition region associated with the unperturbed part $H^0(\zeta_k)$, the LS equation is expressed in the discretized form as
\begin{eqnarray}
\label{eqn:ls05}
\Psi(\zeta_k) &=& \Psi^0(\zeta_k)+{\displaystyle{\sum_{l,l^\prime=1}^m}}{G}^{r0}_T(\zeta_k,\zeta_l)\delta V(\zeta_l,\zeta_{l^\prime})\Psi(\zeta_{l^\prime}) \quad (k=0,1,\cdots,m+1),
\end{eqnarray}
which is referred to as the grid LS equation.
This discretized form within the framework of the RSFD approach unifies Eq.~(\ref{eqn:ls01}) and the scattering boundary conditions. The boundary condition Eq.~(\ref{eqn:ls003}) now reads as
\begin{eqnarray}
\label{eqn:ls06}
\Psi(\zeta_k) &=& 
\left\{
 \begin{array}{ll}
  \Phi^{in}(\zeta_k)+ {\displaystyle{\sum_{j=1}^{N}}}r_j\Phi^{ref}_j(\zeta_k) &$ in the left electrode $ \quad (k\leq 0) \\
  {\displaystyle{\sum_{j=1}^{N}}}t_j\Phi^{tra}_j(\zeta_k) &$ in the right electrode $ \quad (k\geq m+1)
 \end{array}
\right. .
\end{eqnarray}

As the Hamiltonian matrix of the unperturbed reference system, $\hat{H}^0$, is a block-tridiagonal form, the $N$-dimensional block matrix ${G}^{r0}_T (\zeta_k,\zeta_l;E)$, which is a component of the retarded Green's function matrix $\hat{G}^{r0}_T=(E-\hat{H}^0)^{-1}$, is expressed in terms of the scattering wave functions in a variable-separable form as (see Appendix~A) 
\begin{eqnarray}
\label{eqn:ls12}
{G}^{r0}_T(\zeta_k,\zeta_l;E) &=& \left\{
\begin{array}{ll}
U^0_R(\zeta_k){U^0_R(\zeta_l)}^{-1} {D^0_l} & \quad (k < l) \\
{D^0_l} & \quad (k = l) \\
U^0_L(\zeta_k){U^0_L(\zeta_l)}^{-1} {D^0_l} & \quad (k > l) 
\end{array}
\right. .
\end{eqnarray}
Here, $U^0_R(\zeta_k)$ ($U^0_L(\zeta_k)$) is the $N$-dimensional matrix made of the solutions of the Kohn--Sham equation in the case of electrons coming from the right~(left) electrode in the reference system, that is
\begin{eqnarray}
\label{eqn:ls13}
U^0_R(\zeta_k) & = & \Bigl( \Psi^0_{R,1}(\zeta_k),\Psi^0_{R,2}(\zeta_k),\cdots,\Psi^0_{R,N}(\zeta_k) \Bigr), \\
\label{eqn:ls13-2}
U^0_L(\zeta_k) & = & \Bigl( \Psi^0_{L,1}(\zeta_k),\Psi^0_{L,2}(\zeta_k),\cdots,\Psi^0_{L,N}(\zeta_k) \Bigr) ,
\end{eqnarray}
where the $N$-dimensional columnar vector $\Psi^0_{R,j}(\zeta_k)$ $\bigl( \Psi^0_{L,j}(\zeta_k) \bigr)$ denotes the scattering wave functions at $\zeta_k$ for the $j$th incident wave $\Phi^{0,in}_{R,j}$ $\bigl( \Phi^{0,in}_{L,j} \bigr)$ incoming from deep inside the right (left) electrode, where the incident wave is considered to include an evanescent wave as well as an ordinary propagating wave; more precisely, $\left\{\Phi^{0,in}_{R,j}\right\}$ $\bigl( \left\{\Phi^{0,in}_{L,j}\right\} \bigr)$ is taken to be a set of the $N$ generalized Bloch states consisting of leftward (rightward) propagating Bloch waves and decaying evanescent waves toward the left (right) side, which are the solutions of the $2N$-dimensional generalized eigenvalue equation~\cite{fujimoto,icp,onoNEGF}. The matrix $D^0_k$ stands for the diagonal block-matrix element of the retarded Green's function matrix, ${G}^{r0}_T(\zeta_k,\zeta_k;E)$, the representation of which is derived in Appendix~\ref{sec:appA} as (see Eq.~(\ref{eqn:ls75}))
\begin{eqnarray}
\label{eqn:ls14}
D^0_k & = & \bigl[-B^0(\zeta_{k-1})^{\dagger} U^0_R(\zeta_{k-1})U^0_R(\zeta_{k})^{-1} +A^0(\zeta_k) -B^0(\zeta_{k}) U^0_L(\zeta_{k+1})U^0_L(\zeta_{k})^{-1}\bigr]^{-1}
\end{eqnarray}
with $A^0(\zeta_k)$ $\bigl( -B^0(\zeta_k) \bigr)$ being the diagonal (off-diagonal) block-matrix element of $(E-\hat{H}^0)$.

Since $U^0_{R(L)}(\zeta_k)$ includes the exponentially growing or decaying evanescent waves, the calculation using Eq.~(\ref{eqn:ls12}) frequently gives rise to the serious numerical errors~\cite{tsukaLS}. We provide a remedy for this problem as follows.

Introducing the {\it ratio matrices} $X^0_k$ and $Y^0_k$ at {\it two successive} $\zeta_k$ points, which are defined as
\begin{eqnarray}
\label{eqn:ls10-01}
X^0_k &\equiv& U^0_{R}(\zeta_{k-1}) \bigl( U^0_{R}(\zeta_k) \bigr)^{-1} \\
\label{eqn:ls10-02}
Y^0_k &\equiv& U^0_{L}(\zeta_{k+1}) \bigl( U^0_{L}(\zeta_k) \bigr)^{-1},
\end{eqnarray}
respectively, we obtain the following $(m+2)$-dimensional block-matrix expression for the retarded Green's function Eq.~(\ref{eqn:ls12}):
\begin{eqnarray}
\hat{G}^{r0}_T &=&
\left[
\begin{array}{cccccc}
D^0_0 & X^0_1 D^0_1 & X^0_1 X^0_2 D^0_2 & \cdots & {\displaystyle \prod^{m+1}_{j=1}} X^0_j \!\cdot\! D^0_{m+1} \\
Y^0_0 D^0_0 & D^0_1 & X^0_2 D^0_2 & \cdots & {\displaystyle \prod^{m+1}_{j=2}} X^0_j \!\cdot\! D^0_{m+1} \\
\vdots & \vdots& \vdots & \ddots & \vdots \\
{\displaystyle \prod^0_{j=m}} Y^0_j \!\cdot\! D^0_0 \ \ & {\displaystyle \prod^1_{j=m}} Y^0_j \!\cdot\! D^0_1 \ \ & {\displaystyle \prod^2_{j=m}} Y^0_j \!\cdot\! D^0_2 & \cdots & D^0_{m+1}
\end{array}
\right], 
\end{eqnarray}
that is, we rewrite the block-matrix element of $\hat{G}^{r0}_T$ in Eq.~(\ref{eqn:ls12}) as
\begin{eqnarray}
{G}^{r0}_T(\zeta_k,\zeta_l;E) &=& \left\{
\begin{array}{ll}
 {\displaystyle \prod^l_{j=k+1}}\!\! X^0_j \!\cdot\! D^0_l  & \quad (k < l) \\
D^0_l & \quad (k=l) \\
{\displaystyle \prod^l_{j=k-1}}\!\! Y^0_j \!\cdot\! D^0_l  & \quad (k > l)
\end{array}
\right. ,
\label{eqn:ls99}
\end{eqnarray}
and from Eqs.~(\ref{eqn:ls14})--(\ref{eqn:ls10-02}) the diagonal block-matrix element $D^0_k$ reads as
\begin{eqnarray}
D^0_k &=& \bigl[ -B^0(\zeta_{k-1})^{\dagger}X^0_{k} + A^0(\zeta_{k}) -B^0(\zeta_{k})Y^0_k \bigr]^{-1}.
\label{eqn:ls555}
\end{eqnarray}
In the following subsections \ref{ssec:jelele} and \ref{ssec:cryele}, we will give efficient numerical calculation techniques for the ratio matrices $\{X^0_k\}$ and $\{Y^0_k\}$ without employing the matrices $\{U^0_R(\zeta_k)\}$ and $\{U^0_L(\zeta_k)\}$ which include evanescent waves explicitly.
Our previous study~\cite{tsukaLS} verified that the introduction of the ratio expression such as Eqs.~(\ref{eqn:ls10-01})--(\ref{eqn:ls99}) into the retarded Green's function enables us to avoid the numerical collapse originated from the appearance of the rapidly growing and decaying evanescent waves. By contrast, in LS simulations of electron transport through {\it long} conductor systems using the conventional Green's function in a variable-separable form, the numerical collapse is inevitable.

In the solving of Eq.~(\ref{eqn:ls05}) by using the iterative method such as the conjugate gradient method, the operation of $\sum {G}^{r0}_T(\zeta_{k},\zeta_{l})\delta V(\zeta_{l},\zeta_{l^\prime})\Psi(\zeta_{l^\prime})$ in Eq.~(\ref{eqn:ls05}) is carried out as follows:
\begin{eqnarray}
\label{eqn:ls15}
{\sum_{l,l^\prime=1}^m}{G}^{r0}_T(\zeta_k,\zeta_l)\delta V(\zeta_l,\zeta_{l^\prime})\Psi(\zeta_{l^\prime}) &=& \Psi^\prime(\zeta_k)+P_L(\zeta_k)+P_R(\zeta_k) \nonumber \\
& & \quad (k=0,1,\cdots,m+1),
\end{eqnarray}
where
\begin{eqnarray}
\label{eqn:ls16}
\Psi^\prime(\zeta_k) &=& D^0_k{\displaystyle{\sum^m_{l=1}}}\delta V(\zeta_k,\zeta_l)\Psi(\zeta_l), \\
\label{eqn:ls16-2}
P_L(\zeta_k) &=& {\displaystyle \prod^{0}_{j=k-1}}\!Y^0_j \!\cdot\! \Psi^\prime(\zeta_0) + {\displaystyle \prod^{1}_{j=k-1}}\!Y^0_j \!\cdot\! \Psi^\prime(\zeta_1) + \cdots + Y^0_{k-1}\Psi^\prime(\zeta_{k-1}), \\
\label{eqn:ls16-3}
P_R(\zeta_k) &=& X^0_{k+1}\Psi^\prime(\zeta_{k+1}) + X^0_{k+1}X^0_{k+2}\Psi^\prime(\zeta_{k+2}) + \cdots + {\displaystyle \prod^{m+1}_{j=k+1}}\!X^0_j \!\cdot\! \Psi^\prime(\zeta_{m+1}).
\end{eqnarray}
It is easily shown that the sequences $\{ P_L(\zeta_k) \}$ and $\{ P_R(\zeta_k) \}$ satisfy the following recursive relations:
\begin{eqnarray}
\label{eqn:ls18}
P_L(\zeta_k)&=& \left\{
\begin{array}{ll}
0 & \quad (k=0,1) \\
Y^0_{k-1}[P_L(\zeta_{k-1})+\Psi^\prime(\zeta_{k-1})] & \quad (k=2,3,\cdots,m+1)
\end{array}
\right., \\
\label{eqn:ls19}
P_R(\zeta_k)&=& \left\{
\begin{array}{ll}
0 & \quad (k={m+1},m) \\
X^0_{k+1}[P_R(\zeta_{k+1})+\Psi^\prime(\zeta_{k+1})] & \quad (k=m-1,\cdots,1,0)
\end{array}
\right. .
\end{eqnarray}
Here, we used the fact that $\delta V(\zeta_k, \zeta_l)=0$ outside the transition region of $\zeta_1 \leq \zeta_{k(l)} \leq \zeta_m$. 
It should be emphasized that since the elements of $\hat{G}^{r0}_T$ in Eq.~(\ref{eqn:ls99}) are no longer in a variable-separable form, the amount of $[(m+2)N]^2$ for each multiplication is expected to be required; nevertheless, it is reduced to the order of $(m + 2)N^2$ by virtue of Eqs.~(\ref{eqn:ls18}) and (\ref{eqn:ls19}), which means that the present method does not suffer from the numerical collapse without increasing the computational cost.

\subsection{Jellium Electrodes\label{ssec:jelele}}
The case in which electrodes are approximated by structureless jellium models is treated. The jellium electrode approximation has been successfully applied to the interpretation of electron-transport properties with less computational load~\cite{ndlang,nkobayashi,mokamoto,RMNieminen,LS_tsuka1,sfuruya,khirose,tono}. A free electron system is chosen as the unperturbed one with the Hamiltonian $H^0$ where a completely flat potential is assumed, for simplicity. The Green's function in the free-electron system is more conveniently described by using $z_k$ instead of $\zeta_k$. In Appendix~\ref{sec:appB}, we discuss the analytical expression of the Green's function in terms of $z_k$ in a general $\mathcal{N}_f$ case.

We here give details on the implementation of the analytically expressed retarded Green's function in the 3-dimensional central finite-difference ($\mathcal{N}_f$=1) case, for example, which is written by
\begin{eqnarray}
\label{eqn:ls07}
{G}^{r0}_T(\vecvar{r}_{\dsla,j},z_k,\vecvar{r}_{\dsla,j^\prime},z_l;E) &=& \frac{h_z^2}{i{N}} \sum^{\frac{N_x-1}{2}}_{n_x=-\frac{N_x-1}{2}} \sum^{\frac{N_y-1}{2}}_{n_y=-\frac{N_y-1}{2}} \frac{\exp\bigl(i{(\vecvar{G}_{\dsla,n}+\vecvar{k}^B_{\dsla})\cdot(\vecvar{r}_{\dsla,j}-\vecvar{r}_{\dsla,j^\prime})}+i\mathcal{K}_1|z_k-z_l|\bigr)}{\sin{\mathcal{K}_1h_z}}. \nonumber  \\
\end{eqnarray}
Here, 
\begin{eqnarray}
\label{eqn:ls11}
{\vecvar{G}_{\dsla,n}} &=& \left( G_{n_x},G_{n_y} \right) \ =\  \left( \frac{2\pi}{h_xN_x}n_x,\frac{2\pi}{h_yN_y}n_y \right),
\end{eqnarray}
$\vecvar{r}_{\dsla,j}=(x_{j_x}, y_{j_y})$ are the lateral coordinates with $j_{x(y)}=1,2,\cdots,N_{x(y)}$ [$N_{x(y)}$ is chosen an odd integer for convenience], and
\begin{eqnarray}
\label{eqn:ls09-01}
\mathcal{K}_1 &=& \left\{
\begin{array}{lcllcl}
\displaystyle{\frac{1}{h_z}_{}\cos^{-1}\left(1-h_z^2\bigl(E-E_{n_x,n_y}^{(\mathcal{N}_f=1)}\bigr)\right)} & \cdots & E_{n_x,n_y}^{(\mathcal{N}_f=1)} \le E < E_{n_x,n_y}^{(\mathcal{N}_f=1)} + \displaystyle{\frac{2}{h_z^2}} \\
i\displaystyle{\frac{1}{h_z}_{}\cosh^{-1}\left( 1-h_z^2\bigl(E-E_{n_x,n_y}^{(\mathcal{N}_f=1)}\bigr)\right)} & \cdots & E < E_{n_x,n_y}^{(\mathcal{N}_f=1)} \\ 
\displaystyle{\frac{1}{h_z}_{} \left[ \pi+i\cosh^{-1}\left(-1+h_z^2\bigl(E-E_{n_x,n_y}^{(\mathcal{N}_f=1)}\bigr)\right) \right]} & \cdots & E_{n_x,n_y}^{(\mathcal{N}_f=1)} + \displaystyle{\frac{2}{h_z^2}}  \le  E
\end{array}
\right. .
\end{eqnarray}
with 
\begin{eqnarray}
\label{eqn:ls09-02}
E_{n_x,n_y}^{(\mathcal{N}_f=1)} &=& \frac{1}{h_x^2}\left\{1-\cos{\left(G_{n_x}+k^B_x\right) h_x}\right\} + \frac{1}{h_y^2}\left\{1-\cos{\left(G_{n_y}+k^B_y\right) h_y}\right\},
\end{eqnarray}
In the derivation of Eqs.~(\ref{eqn:ls07})--(\ref{eqn:ls09-02}), we used Eqs.~(\ref{eqn:lsB-21}) and (\ref{eqn:lsB-22}) and the extension of Eq.~(\ref{eqn:lsB-26}) to the case of the 3-dimensional space.

Since $D^0_k$ defined by Eq.~(\ref{eqn:ls99}) is the diagonal block-matrix element of the retarded Green's function ${G}^{r0}_T(z_k,z_l;E)$, the $j$th row and $j^\prime$th column element $( D^0_k)_{j,j^\prime}$ is expressed as 
\begin{eqnarray}
\label{eqn:ls21}
(D^0_k)_{j,j^\prime} &\equiv& {G}^{r0}_T(\vecvar{r}_{\dsla,j},z_k,\vecvar{r}_{\dsla,j^\prime},z_k;E) \nonumber \\
&=&\frac{h_z^2}{iN} \sum^{\frac{N_x-1}{2}}_{n_x=-\frac{N_x-1}{2}} \sum^{\frac{N_y-1}{2}}_{n_y=-\frac{N_y-1}{2}} \exp\bigl(i(\vecvar{G}_{\dsla,n}+\vecvar{k}^B_{\dsla})\cdot(\vecvar{r}_{\dsla,j}-\vecvar{r}_{\dsla,j^\prime})\bigr) \times \frac{1}{\sin{\mathcal{K}_1 h_z}}.
\end{eqnarray}
One can see from Eq.~(\ref{eqn:ls21}) that $( D^0_k)_{j,j^\prime}$, and thus $D^0_k$, is $k$-independent owing to the translation invariance in the $z$ direction.
On the other hand, by Eq.~(\ref{eqn:ls99}), $X^0_k$ and $Y^0_k$ are given by
\begin{eqnarray}
X^0_k &=& {G}^{r0}_T(z_{k-1},z_k;E) \left(D^0_{k}\right)^{-1}, \\
Y^0_k &=& {G}^{r0}_T(z_{k+1},z_k;E) \left(D^0_{k}\right)^{-1},
\end{eqnarray}
and from Eq.~(\ref{eqn:ls07}), the $j$th row and $j^\prime$th column matrix element of ${G}^{r0}_T(z_{k \pm 1},z_k;E)$ are described as
\begin{eqnarray}
\label{eqn:ls22}
\left({G}^{r0}_T(z_{k \pm 1},z_k;E)\right)_{j,j^\prime}&\equiv&{G}^{r0}_T(\vecvar{r}_{\dsla,j},z_{k \pm 1},\vecvar{r}_{\dsla,j^\prime},z_k;E) \nonumber \\
&=& \frac{h_z^2}{iN} \!\!\! \sum^{\frac{N_x-1}{2}}_{n_x=-\frac{N_x-1}{2}} \sum^{\frac{N_y-1}{2}}_{n_y=-\frac{N_y-1}{2}} \!\!\! \exp\bigl(i(\vecvar{G}_{\dsla,n}+\vecvar{k}^B_{\dsla})\cdot(\vecvar{r}_{\dsla,j}-\vecvar{r}_{\dsla,j^\prime})\bigr) \frac{1}{\sin{\mathcal{K}_1 h_z}} \exp\bigl(i\mathcal{K}_1 h_z\bigr). \nonumber \\
\end{eqnarray}
This implies that $X^0_k$ and $Y^0_k$ are also $k$-independent and
\begin{eqnarray}
\label{eqn:ls22-2}
X^0_k &=& Y^0_k.
\end{eqnarray}
After some calculations,
\begin{eqnarray}
\label{eqn:ls22-3}
(X^0_k)_{j,j^\prime}&\equiv& \left[ {G}^{r0}_T(z_{k-1},z_k;E)(D^0_k)^{-1} \right]_{j,j^\prime} \nonumber \\
&=& \frac{1}{N} \!\!\! \sum^{\frac{N_x-1}{2}}_{n_x=-\frac{N_x-1}{2}} \sum^{\frac{N_y-1}{2}}_{n_y=-\frac{N_y-1}{2}} \!\!\! \exp\bigl(i(\vecvar{G}_{\dsla,n}+\vecvar{k}^B_{\dsla})\cdot(\vecvar{r}_{\dsla,j}-\vecvar{r}_{\dsla,j^\prime})\bigr) \times \exp\bigl(i\mathcal{K}_1 h_z\bigr)
\end{eqnarray}
is obtained~\cite{comm1}.
Hereafter, $D^0_k$ and $X^0_k$ are denoted by $D^0$ and $X^0$, respectively, since they are $k$-independent.

The products of the matrix $D^0$ $\left( X^0 \right)$ and vectors $\left\{f(\vecvar{r}_{\dsla,j},z_k)|~j=1,2,\cdots,N\right\}$ as required in the computations of Eqs.~(\ref{eqn:ls16}), (\ref{eqn:ls18}) and (\ref{eqn:ls19}) can be easily carried out in the momentum space, since they are written in the convolution form of the 2-dimensional discrete Fourier transform. Owing to the orthogonality of the plane waves, the Fourier transformed $D^0$ and $X^0$ are represented as the diagonalized matrices, i.e.,
\begin{eqnarray}
\label{eqn18-2}
\mathcal{F}\left[ D^0 \right]_{n,n^\prime} &\equiv& \frac{1}{N} \sum_{\vecvar{r}_{\dsla,j}} \sum_{\vecvar{r}_{\dsla,j^\prime}} \exp\bigl(-i(\vecvar{G}_{\dsla,n}+\vecvar{k}^B_{\dsla})\cdot\vecvar{r}_{\dsla,j}+i(\vecvar{G}_{\dsla,n^\prime}+\vecvar{k}^B_{\dsla})\cdot\vecvar{r}_{\dsla,j^\prime}\bigr) (D^0)_{j,j^\prime} \nonumber \\
&=& -ih_z^2 \delta_{nn^\prime} \frac{1}{\sin(\mathcal{K}_1h_z)},\\
\label{eqn18-2-2}
\mathcal{F}\left[ X^0 \right]_{n,n^\prime} &\equiv& \frac{1}{N} \sum_{\vecvar{r}_{\dsla,j}} \sum_{\vecvar{r}_{\dsla,j^\prime}} \exp\bigl(-i(\vecvar{G}_{\dsla,n}+\vecvar{k}^B_{\dsla})\cdot\vecvar{r}_{\dsla,j}+i(\vecvar{G}_{\dsla,n^\prime}+\vecvar{k}^B_{\dsla})\cdot\vecvar{r}_{\dsla,j^\prime}\bigr) (X^0)_{j,j^\prime} \nonumber \\
&=& \delta_{nn^\prime} \exp(i\mathcal{K}_1h_z),
\end{eqnarray}
respectively. Finally, one can obtain the matrix elements of the Fourier transform of the terms shown in Eqs.~(\ref{eqn:ls16}), (\ref{eqn:ls18}) and (\ref{eqn:ls19}) as
\begin{eqnarray}
\label{eqn:ls24}
\mathcal{F} \bigl[ \Psi^\prime(z_k) \bigr]_{n} \!&=&\! \sum_{n^\prime} \mathcal{F}\left[ D^0 \right]_{n,n^\prime} \mathcal{F} \left[ {\displaystyle{\sum_l}}\delta V(z_k,z_l)\Psi(z_l) \right]_{n^\prime}, \\
\label{eqn:ls25}
\mathcal{F} \bigl[ P_L(z_k) \bigr]_{n} \!&=&\! \left\{
\begin{array}{ll}
0 & \quad (k\!=\!0,1) \\
\displaystyle{ \sum_{n^\prime}} \mathcal{F}\left[X^0\right]_{n,n^\prime} \bigl( \mathcal{F} \bigl[ P_L(z_{k-1})\bigr]_{n^\prime}+\mathcal{F} \bigl[ \Psi^\prime(z_{k-1}) \bigr]_{n^\prime} \bigr) & \quad (k\!=\!2,3,\cdots,m\!+\!1)
\end{array}
\right., \nonumber \\ \\
\label{eqn:ls27}
\mathcal{F} \bigl[ P_R(z_k) \bigr]_{n} \!&=&\! \left\{
\begin{array}{ll}
0 & \quad (k\!=\!m\!+\!1,m) \\
\displaystyle{ \sum_{n^\prime}} \mathcal{F}\left[X^0\right]_{n,n^\prime} \bigl( \mathcal{F} \bigl[ P_R(z_{k+1}) \bigr]_{n^\prime}+\mathcal{F} \bigl[ \Psi^\prime(z_{k+1}) \bigr]_{n^\prime} \bigr) & \quad (k\!=\!m\!-\!1,\cdots,1,0)
\end{array}
\right. , \nonumber \\ 
\end{eqnarray}
respectively.

For calculating the product in Eq.~(\ref{eqn:ls05}), the computational cost of $O(N_{in} \times N_z^2 \times N^2)$ is required.
However, by introducing the 2-dimensional discrete fast Fourier transform~(FFT) algorithm, the cost of the product shown in Eqs.~(\ref{eqn:ls24})--(\ref{eqn:ls27}) decreases to $O(N_{in} \times N_z \times N\log{N})$, since the off-diagonal elements of the Fourier transformed matrices $\mathcal{F}\left[ D^0 \right]$ and $\mathcal{F}\left[X^0\right]$ are zero as seen in Eqs.~(\ref{eqn18-2}) and (\ref{eqn18-2-2}).
The Fourier transform of a columnar vector and the inverse Fourier transform of $\mathcal{F}\left[ D^0 \right] \times \mathcal{F}\left[ f_k \right]$ and $\mathcal{F}\left[ X^0 \right] \times \mathcal{F}\left[ f_k \right]$ are carried out at each $z_k$ point using FFT algorithm. Here, $\mathcal{F}\left[ f_k \right]$ represents the Fourier transformed vector of $\left\{f(\vecvar{r}_{\dsla,j},z_k)|~j=1,2,\cdots,N\right\}$. Thus, the maximum order of the calculations is improved from $O(N_{in} \times N_z^2 \times N^2)$ to $O(N_{in} \times N_z \times N\log N)$.
The above mentioned discussion on the central finite-difference approximation can be straightforwardly extended to the cases of the higher-order finite-difference approach.

\subsection{Crystalline Electrodes\label{ssec:cryele}}
A general case is discussed where a system with atomistic crystalline electrodes is chosen as the unperturbed reference system; one electrode is confronted with the other across the empty transition region. We present efficient procedures for calculating the ratio matrices $X^0_k$ and $Y^0_k$ in this case.

The matrices $X^0_0$ and $Y^0_{m+1}$ defined by Eqs.~(\ref{eqn:ls10-01}) and (\ref{eqn:ls10-02}) are described as
\begin{eqnarray}
\label{egami:eqn9-122}
X^0_0 & \equiv & U^0_R(\zeta_{-1})\bigl( U^0_R(\zeta_{0}) \bigr)^{-1} \hspace{6.5mm} = \bigl( B^0(\zeta_{-1})^{\dagger} \bigr)^{-1}{\textstyle \sum}^{r0}_L(\zeta_0), \\ 
\label{egami:eqn9-123}
Y^0_{m+1} & \equiv & U^0_L(\zeta_{m+2})\bigl( U^0_L(\zeta_{m+1}) \bigr)^{-1} = \bigl( B^0(\zeta_{m+1}) \bigr)^{-1}{\textstyle \sum}^{r0}_R(\zeta_{m+1}),
\end{eqnarray}
where $\sum^{r0}_L(\zeta_0)\ \bigl(\sum^{r0}_R(\zeta_{m+1})\bigr)$ is the self-energy term defined on the left- (right-)electrode surface and can be calculated by using the continued-fraction equation; for the details of the derivation of Eqs.~(\ref{egami:eqn9-122}) and (\ref{egami:eqn9-123}) and the computation of the self-energy terms, see Refs.~10 and 18. %Refs.~\cite{icp} and \cite{onoNEGF}.
For the sake of comparison, we note that $U_{R(L)}(\zeta_k)$ defined by Eqs.~(\ref{eqn:ls13}) and (\ref{eqn:ls13-2}) is identical to $Q^{p(q)}(\zeta_k)$ of Eq.~(15) in Ref.~18. We also emphasize that the accuracy of $\Sigma_{R(L)}(\zeta_k)$ is enhanced by making use of the continued-fraction equation in a self-consistent manner, as shown in Eqs.~(16)--(18) in Ref.~18.%Ref.~18=\cite{onoNEGF}

It should be noticed that the terms $\{ X^0_k \}$ can be sequentially computed as
\begin{eqnarray}
X^0_1 &=& \bigl(A^0(\zeta_0)-B^0(\zeta_{-1})^{\dagger}X^0_0 \bigr)^{-1}B^0(\zeta_0) \nonumber \\
X^0_2 &=& \bigl(A^0(\zeta_1)-B^0(\zeta_{0})^{\dagger}X^0_1 \bigr)^{-1}B^0(\zeta_1) \nonumber \\
X^0_3 &=& \bigl(A^0(\zeta_2)-B^0(\zeta_{1})^{\dagger}X^0_2 \bigr)^{-1}B^0(\zeta_2) \nonumber \\
& \vdots& \nonumber \\
X^0_{m+1} &=& \bigl( A^0(\zeta_m)-B^0(\zeta_{m-1})^{\dagger}X^0_m \bigr)^{-1}B^0(\zeta_m),
\label{egami:eqn9-134}
\end{eqnarray}
which are easily derived from Eq.~(\ref{eqn:ls671}), and similarly, the iterative series of $\{ Y^0_k \}$ are obtainable from Eq.~(\ref{eqn:ls672}) as
\begin{eqnarray}
Y^0_m     &=& \bigl(A^0(\zeta_{m+1})-B^0(\zeta_{m+1})Y^0_{m+1} \bigr)^{-1}B^0(\zeta_m)^{\dagger} \nonumber \\
Y^0_{m-1} &=& \bigl(A^0(\zeta_{m})-B^0(\zeta_{m})Y^0_{m} \bigr)^{-1}B^0(\zeta_{m-1})^{\dagger} \nonumber \\
Y^0_{m-2} &=& \bigl(A^0(\zeta_{m-1})-B^0(\zeta_{m-1})Y^0_{m-1} \bigr)^{-1}B^0(\zeta_{m-2})^{\dagger} \nonumber \\
&  \vdots& \nonumber \\
Y^0_{0}   &=& \bigl(A^0(\zeta_{1})-B^0(\zeta_{1})Y^0_{1} \bigr)^{-1}B^0(\zeta_{0})^{\dagger}.
\label{egami:eqn9-135}
\end{eqnarray}

The recursive relations Eqs.~(\ref{egami:eqn9-134}) and (\ref{egami:eqn9-135}) allow us to calculate all the matrix elements by a linear scaling operation (order-$N$ calculation procedure) at a limited computational cost. It is also noted that using Eqs.~(\ref{egami:eqn9-134}) and (\ref{egami:eqn9-135}), $X^0_k$ and $Y^0_k$ are stably computed without involving error accumulation since the errors due to the appearance of evanescent waves are eliminated by introducing the ratios of these waves at two successive grid points. Finally, the diagonal block-matrix element $D_k^0$ is given by Eq.~(\ref{eqn:ls555}). Once $D^0_k$, $X^0_k$ and $Y^0_k$ for any $k$ ($0 \le k \le m+1 $) are known, all of the matrix elements of $\hat{G}^{r0}_T$ in Eq.~(\ref{eqn:ls99}) are determined, and the algorithm of Eqs.~(\ref{eqn:ls18}) and (\ref{eqn:ls19}) can be utilized.

\newpage
%!!!!!!!!!!!!!!!!!!!!!!!!!!!!!!!!!!!!!!!!!!!!!!!!!!!!!!!!!!!!!!!!!!!!
\section{\label{sec:level3}Applications}
To demonstrate the performance of the grid LS method, we examine the electron-transport properties of models of semiconductor/insulator interfaces sandwiched between semi-infinite electrodes. Recently, the germanium-based metal-oxide-semiconductor field-effect transistor has attracted significant attention because the electronic band gap of germanium ($\sim0.66$~eV) is lower than that of silicon ($\sim1.12$~eV), which allows for reduced operating voltages. In a highly integrated circuit, it is known that a large leakage current is induced by defects such as impurities and oxygen vacancies in the thin gate oxide layer. So, the relationship between the DB introduced by defects and leakage current in Si/SiO$_2$ interfaces has been extensively investigated~\cite{kageSi,HoussaSi1,YangSi,BinderSi,BroqvistSi,HoussaSi2,TsetserisSi,onoSi}, while the role of the DB state in Ge/GeO$_2$ interfaces is controversial.
One of the present authors~(T.~O.) has performed several investigations on Ge/GeO$_2$ interfaces~\cite{saitoGe1,saitoGe2,onoGe}. In recent work, the relationship between atomic configurations and electronic structures of (001)Si/SiO$_2$ and (001)Ge/GeO$_2$ models with DBs was explored using first-principles simulations within the framework of the local density approximation~(LDA)\cite{lda}. It was found that the Si-DB state is located near the midgap of the Si substrate corresponding to the Fermi level, while the Ge-DB state lies near the top of the valence band which is 0.3~eV below the Fermi level\cite{onoGe}.

To examine how DB states with different characteristics affect leakage currents, we performed transport simulations of electrons flowing across the (001)Si/SiO$_2$ and (001)Ge/GeO$_2$ models. 
The magnitude of the leakage current flowing through insulators is so small that it can be easily affected by interactions between electrodes and interface models and by the value of the energy band gap, which is underestimated by the LDA calculation. Therefore, in this paper, we discuss qualitatively the ratio of the leakage current between models with and without a defect.

\begin{figure}[p]
%\begin{figure}[htb]
\begin{center}
\includegraphics[width=8.0cm]{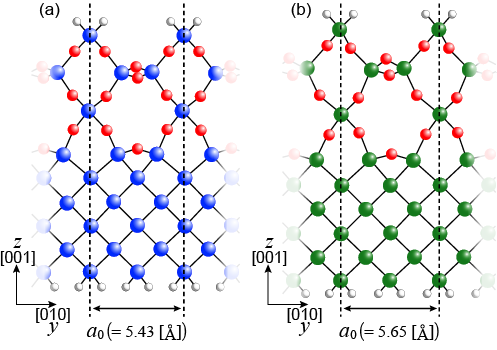}
\caption{(Color online) 
Schematic views of unit cells of (a) Si/SiO$_2$ and (b) Ge/GeO$_2$ models. Dashed lines indicate the boundaries of the cell. White, blue, green, and red spheres represent H, Si, Ge, and O atoms, respectively.
}
\label{fig:SiGe_modelS}
\end{center}
\end{figure}

\begin{figure}[p]
%\begin{figure}[htb]
\begin{center}
\includegraphics*[width=16.0cm]{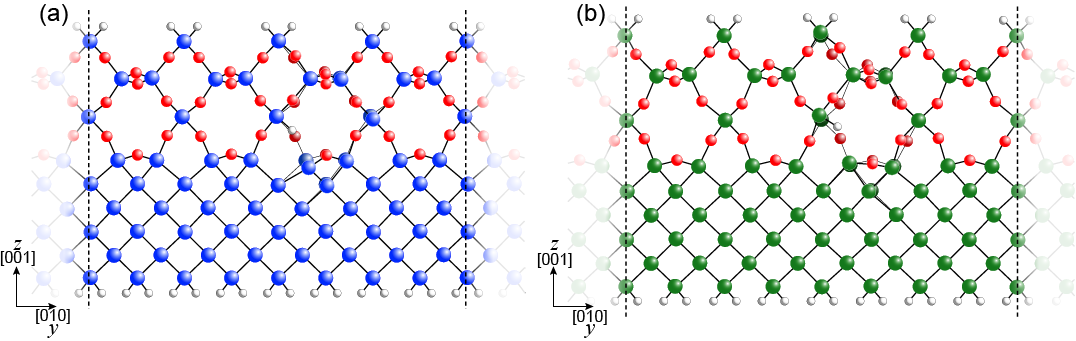}
\caption{(Color online) 
Schematic views of (a) Si/SiO$_2$ and (b) Ge/GeO$_2$ models with an oxygen vacancy after geometrical optimization. Dashed lines indicate the boundaries of supercells. The key to the symbols is the same as in Fig.~\ref{fig:SiGe_modelS}.
}
\label{fig:SiGe_model}
\end{center}
\end{figure}

Figure~\ref{fig:SiGe_modelS} illustrates a unit cell of each interface model. In these models, the side lengths of the cell in the lateral direction parallel to the interface for the Si/SiO$_2$ (Ge/GeO$_2$) model were taken to be the experimental lattice constant of bulk Si~(Ge), $a_0=5.43~(5.65)$~\AA. The thicknesses of the SiO$_2$ (GeO$_2$) layer and the Si (Ge) substrate were 7.34 (7.25) and 7.18 (7.37)~\AA, respectively.
In calculations for the models with an oxygen vacancy, we introduced the defect into a supercell comprising $4 \times 4$ unit cells in the lateral direction (Fig.~\ref{fig:SiGe_model}); this is large enough to avoid interactions between defects in neighboring cells. Two Si~(Ge) DBs were generated near the interface between the Si~(Ge) substrate and the oxide layer in one unit cell by removing a bridging oxygen atom in a manner similar to that used in the previous study\cite{onoGe}. One of the DBs is passivated by a hydrogen atom, while the other remains with the Si (Ge) atom of the center back-bonded to two neighboring Si (Ge) atoms and an oxygen atom $\bigl({}^\bullet$Si$\equiv$Si$_2$O (${}^\bullet$Ge$\equiv$Ge$_2$O)$\bigr)$.
For the no-defect models, the unit cell of each model, depicted in Fig.~\ref{fig:SiGe_modelS}, was employed with $4 \times 4$ sampling $k$-points in the 2-dimensional Brillouin zone for comparison with the models having defects. 

We first optimized the atomic and electronic structures of the models. First-principles calculations based on the RSFD approach were performed in the manner described in Ref.~[42] with a grid spacing of 0.15~\AA.
The size of the supercell in the [001] ($z$) direction was taken to be $5a_0$, including a large enough vacuum region, and the top and bottom layers of the models were terminated by hydrogen atoms.
As shown in Fig.~\ref{fig:SiGe_model}, which illustrates the relaxed configurations, the Si atom with the DB is pulled down to the Si substrate whereas the Ge atom with the DB is slightly raised toward the oxide layer. 

Next, we examined the leakage currents caused by introducing the oxygen vacancies into the models.
Employing the optimized effective Kohn--Sham potential, we used the grid LS method to evaluate the scattering wave functions for electrons incident from the bottom-side electrode. The conductance at the limits of zero temperature and bias are described by the Landauer--B{\"u}ttiker formula~\cite{buttiker}.
In the transport calculation, the top and bottom sides of each model were connected to aluminum jellium electrodes without terminating hydrogen atoms. The Wigner--Seitz radius $r_s$ was $2.07$, which corresponds to the valence electron density of bulk aluminum.

\begin{figure}[p]
%\begin{figure}[htb]
\begin{center}
\includegraphics[width=8.0cm]{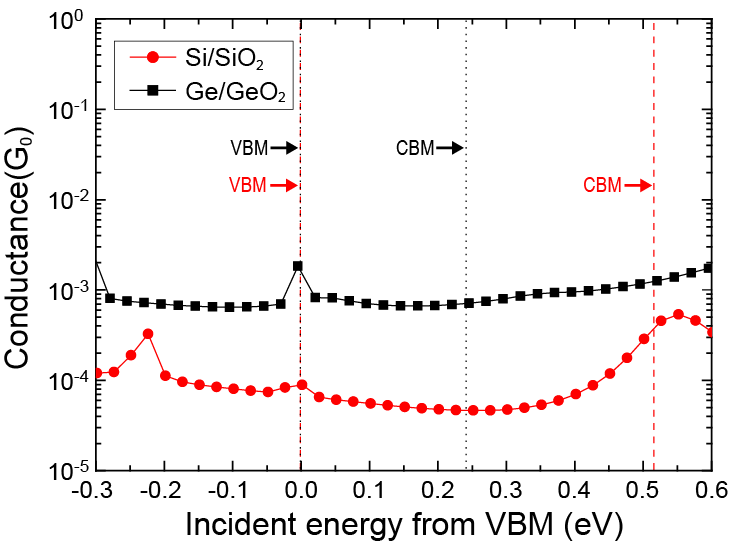}
\caption{(Color online) 
Conductance of Si/SiO$_2$ and Ge/GeO$_2$ models as functions of incident electron energy measured from the valence band maximum (VBM) of the substrates. Red circles and black squares represent the conductance spectra of Si/SiO$_2$ and Ge/GeO$_2$ models, respectively. VBM and conduction band minimum (CBM) of Si (Ge) substrate are indicated by vertical dashed (dotted) lines.
}
\label{fig:SiGe_condS}
\end{center}
\end{figure}

\begin{figure}[p]
%\begin{figure}[htb]
\begin{center}
\includegraphics[width=8.0cm]{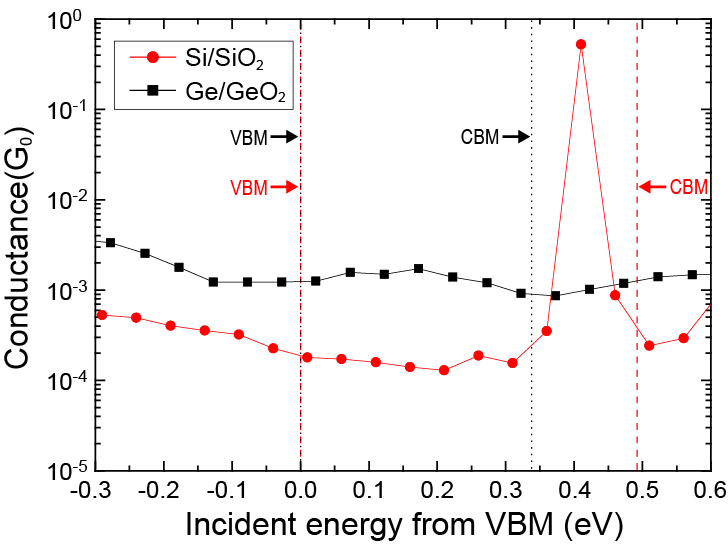}
\caption{(Color online) 
Conductance of Si/SiO$_2$ and Ge/GeO$_2$ models as functions of incident electron energy measured from the valence band maximum (VBM) of the substrates. The key to the symbols is the same as in Fig.~\ref{fig:SiGe_condS}. VBM and conduction band minimum (CBM) of Si (Ge) substrate are indicated by vertical dashed (dotted) lines.
}
\label{fig:SiGe_Cond}
\end{center}
\end{figure}

\begin{figure}[p]
%\begin{figure}[htb]
\begin{center}
\includegraphics[width=8.0cm]{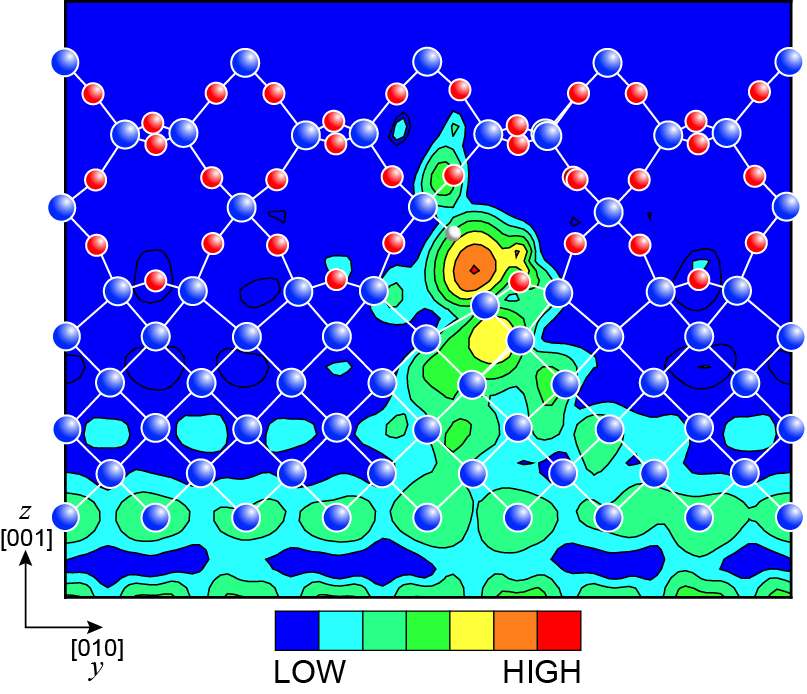}
\caption{(Color online)
Contour plot of the charge-density distribution of electrons flowing over the Si/SiO$_2$ model with incident energy of ${\rm VBM}+0.41$~eV. The key to the symbols is the same as in Fig.~\ref{fig:SiGe_modelS}(a). Here the charge density is integrated in the [100]~($x$) direction. Each contour represents twice or half the density of adjacent contours; the lowest contour is $1.97 \times 10^{-4}$~electron/eV/\AA$^2$.
}
\label{fig:Si_CD}
\end{center}
\end{figure}

\begin{figure}[p]
%\begin{figure}[htb]
\begin{center}
\includegraphics[width=16.0cm]{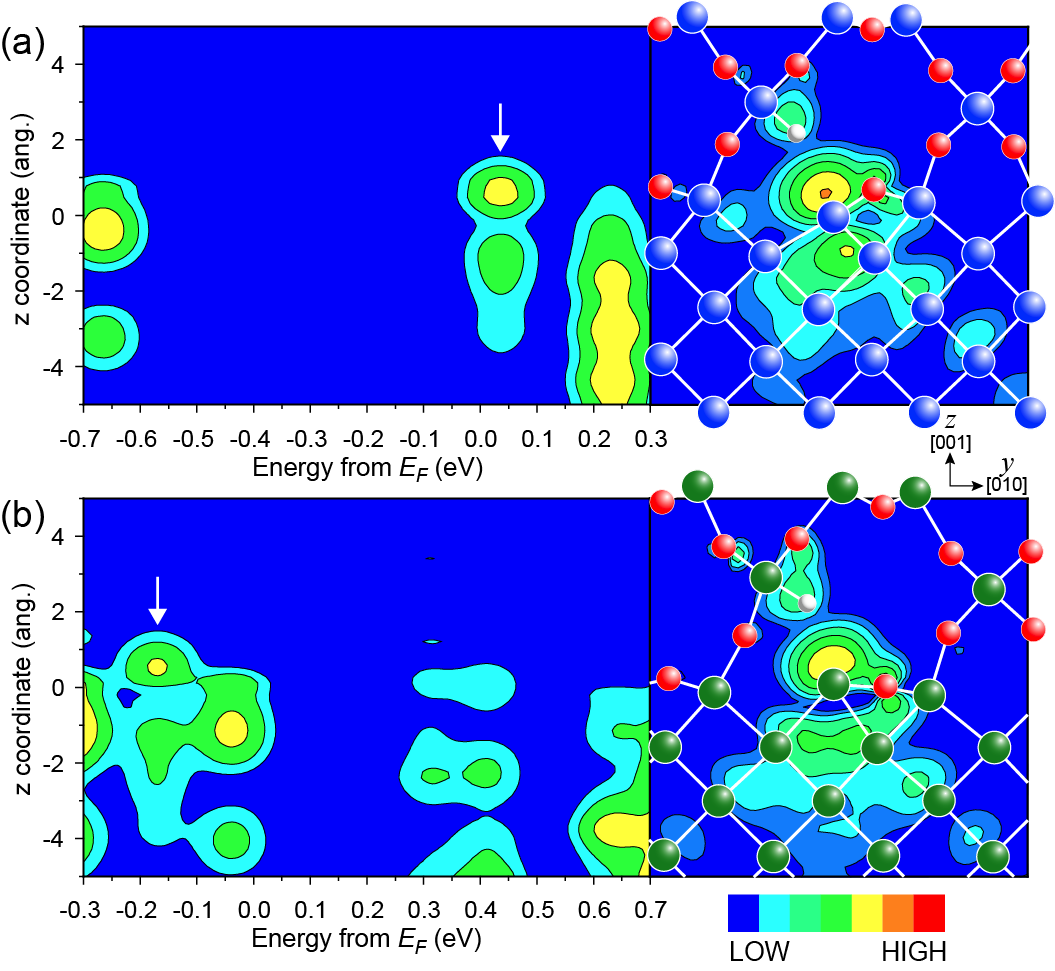}
\caption{(Color online) 
Contour plots of the density of states~(DOS) (left panels) and charge density of dangling-bond (DB) states (right panels) for (a) Si/SiO$_2$ and (b) Ge/GeO$_2$ models without electrodes. Arrows denote the DOS peaks derived from the DB states. Energies are measured from the Fermi level $E_F$ and the $z$ coordinate of the atom with DB is set to zero.
In the right panels, the charge density is integrated in the [100]~($x$) direction.
The key to the symbols in the right panels is the same as in Fig.~\ref{fig:SiGe_modelS}.
The coordinate in the $z$ direction corresponds to that of the left panel. 
Each contour represents twice or half the density of adjacent contours; the lowest contour is $2.56 \times 10^{-2}$~electron/eV/\AA \ ($2.79 \times 10^{-3}$~electron/\AA$^2$) in the left (right) panels.
}
\label{fig:Si_LDOS}
\end{center}
\end{figure}

Figure~\ref{fig:SiGe_condS} shows the computed conductance spectra for the no-defect Si/SiO$_2$ and Ge/GeO$_2$ models as functions of incident electron energy measured from the valence band maximum (VBM) of the substrate. Although some small peaks derived from the bulk states in the valence band of the Si and Ge substrates appear in Fig.~\ref{fig:SiGe_condS}, both models exhibit highly suppressed conductivities in the band-gap region between the VBM and conduction band minimum (CBM) of the substrates. 
Figure~\ref{fig:SiGe_Cond} represents the conductance spectra for the Si/SiO$_2$ and Ge/GeO$_2$ models with the oxygen vacancy. No remarkable peaks appear in the spectrum of the Ge/GeO$_2$ model; however, for the Si/SiO$_2$ model, a peak with high transmission occurs around $\rm{VBM}+0.41$~eV, where electrons flow through the oxide layer via the Si-DB state as shown in the charge density distribution of the scattering electron (Fig.~\ref{fig:Si_CD}). 
In addition, Fig.~\ref{fig:Si_LDOS} exhibits contour plots of the density of states (DOS) integrated in the lateral directions (left panels) and the charge density distributions of the DB states (right panels) for the defect-introduced models without electrodes. In Fig.~\ref{fig:Si_LDOS}(a), the white arrow identifies the peak in the DOS derived from the DB state between the VBM and CBM of the Si substrate. In Fig.~\ref{fig:Si_LDOS}(b), the Ge-DB state is coupled to the states in the valence band of the Ge substrate. The relative positions of the VBM, CBM, and DB state are modulated when each model is connected to electrodes.
When a DB appears in diamond-structured semiconductors, there are two possibilities: the DB state tends to become either more $s$ type or more $p$ type~\cite{hamman}.
In the $p$ type DB state, the three remaining bonds tend to become $sp^2$ hybridized and, to reduce the strain, prefer to be in a plane. This occurs in Fig.~\ref{fig:Si_LDOS}(a) wherein the Si atom with the DB is pulled down and the DB state spatially extends in the [001]~(z) direction. This behavior degrades the insulating properties of the Si/SiO$_2$ model.
In contrast, when the DB state is inclined to be an $s$ type, the three remaining bonds tend to become $p$ types. In this case, the angular separation of these bonds is reduced from that in the tetrahedral structure where the separation angle is 109.5$^\circ$. As a result, the atom with the DB moves away from the three bonded atoms. Therefore, for the DB state of the raised Ge atom, the charge density of the state is distributed in the lateral directions compared with that of the Si-DB state, and the Ge-DB state is coupled with interface states of the Ge substrate~(Fig.~\ref{fig:Si_LDOS}(b)). This behavior barely contributes to electron transport across the model.
Consequently, by introducing the oxygen vacancy, the leakage current in the Si/SiO$_2$ model increases by a factor of 162.9, while that in the Ge/GeO$_2$ model increases by a factor of 11.8~\cite{comment}.

%!!!!!!!!!!!!!!!!!!!!!!!!!!!!!!!!!!!!!!!!!!!!!!!!!!!!!!!!!!!!!!!!!!!!
\section{\label{sec:level4}Conclusion}
We have presented the grid LS equation method based on the fully real-space algorithms to elucidate the scattering wave functions in nanoscale structures sandwiched between semi-infinite electrodes. It is shown that the numerical collapse due to the exponentially growing and decaying evanescent waves and the computational costs can be restrained by using the ratio expression of the retarded Green's function obtained analytically (jellium electrode case) and by incorporating the self-energy matrices and applying the recursive formulas to the ratio matrices (crystalline electrode case).
To demonstrate the performance of our method, we used it to calculate the transport properties of (001)Si/SiO$_2$ and (001)Ge/GeO$_2$ models attached to semi-infinite electrodes. The results show that the DB state in the Ge/GeO$_2$ model gives a much smaller contribution to leakage current than that in the Si/SiO$_2$ model.
Our procedure can precisely and efficiently extend knowledge of the physics underlying the transport of electrons through nanoscale structures.

%!!!!!!!!!!!!!!!!!!!!!!!!!!!!!!!!!!!!!!!!!!!!!!!!!!!!!!!!!!!!!!!!!!!!
\section*{\label{sec:level5}Acknowledgments}
This research was supported by a Grant-in-Aid for Young Scientists (B) Grant No. 24710152 from the Ministry of Education, Culture, Sports, Science and Technology.
The numerical calculation was carried out with the computer facilities at the Institute for Solid State Physics at the University of Tokyo, the Information Synergy Center at Tohoku University.

%!!!!!!!!!!!!!!!!!!!!!!!!!!!!!!!!!!!!!!!!!!!!!!!!!!!!!!!!!!!!!!!!!!!!
\newpage
%!!!!! Appendix !!!!!
\appendix
%%%%%%%%%%%%%%%%%%%%%%%%%%%%%%%%%%%%%%%%%%%%%%%%%%%%%%%%%%%%%%%%%%%%%%%%%%%%%%%
%%%%%%%%%%%%%%%%%%%%%%%%%%%%%%%%%%%%%%%%%%%%%%%%%%%%%%%%%%%%%%%%%%%%%%%%%%%%%%%
\section{Variable-separable-formed retarded Green's function in the RSFD approach\label{sec:appA}}
In this appendix, the subscript $0$ denoting the reference system is omitted, for simplicity.
Let us consider the product of the matrix $(E-\hat{H})$ and the $l$th columnar vector of $\hat{G}^r(E)$, $\bigl\{ G^r(\zeta_k,\zeta_l;E) \bigr\}$ ($k=\cdots,l-1,l,l+1,\cdots$). The retarded Green's function is constructed from {\it outwardly} propagating and decreasing waves, and then, taking it into account, we assume this columnar vector to be represented by
\begin{eqnarray}
\label{eqn:lsA1}
 \Bigl[ \cdots,U_R(\zeta_{l-2})C_R(\zeta_l),U_R(\zeta_{l-1})C_R(\zeta_l),D(\zeta_l),U_L(\zeta_{l+1})C_L(\zeta_l),U_L(\zeta_{l+2})C_L(\zeta_l),\cdots \Bigr]^t &,&
\end{eqnarray}
where $C_{R(L)}(\zeta_l)$ and $D(\zeta_l)$ $\bigl(\equiv G^r(\zeta_l,\zeta_l;E)\bigr)$ are unknown block matrices, and $U_{R(L)}(\zeta_l)$ is defined by Eqs.~(\ref{eqn:ls13}) and (\ref{eqn:ls13-2}). Hereafter, $U_{R(L)}(\zeta_l)$, $C_{R(L)}(\zeta_l)$ and $D(\zeta_l)$ are abbreviated to $U^{R(L)}_l$, $C^{R(L)}_l$ and $D_l$, respectively. By definition, the abovementioned product satisfies
\begin{eqnarray}
\label{eqn:ls66}
\left( 
\begin{array}{lcccccr}
\ddots & \ddots & & & & & \\
\ddots & A_{l-2} & -B_{l-2} & & & \mbox{{\LARGE $0$}}& \\
 & -B_{l-2}^{\dagger} & A_{l-1} & -B_{l-1} & & & \\
 & & -B_{l-1}^{\dagger} & A_{l} & -B_{l} & & \\
 & & & -B_{l}^{\dagger} & A_{l+1} & -B_{l+1} & \\
 &\mbox{{\LARGE $0$}}& & & -B_{l+1}^{\dagger} & A_{l+2} & \ddots \\
 & & & & & \ddots & \ddots
\end{array}
\right) \left( 
\begin{array}{c}
 \vdots \\
 U^R_{l-2}C^R_l \\
 U^R_{l-1}C^R_l \\
 D_l \\
 U^L_{l+1}C^L_l \\
 U^L_{l+2}C^L_l \\
 \vdots
\end{array}
\right) &=& \left( 
\begin{array}{c}
 \vdots \\
 0 \\
 0 \\
 I \\
 0 \\
 0 \\
 \vdots
\end{array}
\right) \leftarrow\mbox{the $l$th}, \ \ \ \ 
\end{eqnarray}
where $A_l=E-H(\zeta_l)$ and $B^{(\dagger)}_l=B^{(\dagger)}(\zeta_l)$.
Since $\bigl\{U^{R(L)}_l\bigr\}$ is a set of the solutions of the Kohn--Sham equation, the following equations hold:
\begin{eqnarray}
\label{eqn:ls671}
-B^{\dagger}_{k-1}U^R_{k-1}+A_kU^R_k-B_{k}U^R_{k+1} &=& 0 , \\
\label{eqn:ls672}
-B^{\dagger}_{k-1}U^L_{k-1}+A_kU^L_k-B_{k}U^L_{k+1} &=& 0 , \\
(k=\cdots,l-1,l,l+1,\cdots) && \nonumber
\end{eqnarray}
From Eqs.~(\ref{eqn:ls66})--(\ref{eqn:ls672}), one sees that the unknown matrices $C_l^{R(L)}$ and $D_l$ are required to satisfy the equations
\begin{eqnarray}
\label{eqn:ls68}
B_{l-1}U^{R}_{l}C^{R}_{l} - B_{l-1}D_{l} &=& 0 ,\\
\label{eqn:ls69}
-B^{\dagger}_{l-1}U^{R}_{l-1}C^{R}_{l} + A_{l}D_{l} - B_{l}U^{L}_{l+1}C^{L}_{l} &=& I,\\
\label{eqn:ls70}
-B^{\dagger}_{l}D_{l} + B^{\dagger}_{l}U^{L}_{l}C^{L}_{l} &=& 0,
\end{eqnarray}
and thus, Eqs.~(\ref{eqn:ls68}) and (\ref{eqn:ls70}) lead to the relationships between $C_l^{R(L)}$ and $D_l$ as 
\begin{eqnarray}
\label{eqn:ls71}
C^{R}_{l} & = & \bigl( U^{R}_{l} \bigr)^{-1}D_{l}, \\
\label{eqn:ls72}
C^{L}_{l} & = & \bigl( U^{L}_{l} \bigr)^{-1}D_{l},
\end{eqnarray}
and Eq.~(\ref{eqn:ls69}) decides $D_l$ to be
\begin{eqnarray}
\label{eqn:ls75}
 D_{l} &=& \Bigl[ -B^{\dagger}_{l-1} U^{R}_{l-1} \bigl( U^{R}_{l} \bigr)^{-1} + A_{l} - B_{l}U^{L}_{l+1} \bigl( U^{L}_{l} \bigr)^{-1} \Bigr]^{-1}.
\end{eqnarray}
In consequence, the retarded Green's function $G^{r}_T(\zeta_k,\zeta_l;E)$ can be described in the following separable form:
\begin{eqnarray}
\label{eqn:ls74}
G^{r}_T(\zeta_k,\zeta_l;E) &=& \left\{
\begin{array}{ll}
U^R_k \bigl( U^R_l \bigr)^{-1} D_l & \quad (k < l) \\
 D_l & \quad (k = l) \\
U^L_k \bigl( U^L_l \bigr)^{-1} D_l & \quad (k > l) 
\end{array}
\right. .
\end{eqnarray}

%%%%%%%%%%%%%%%%%%%%%%%%%%%%%%%%%%%%%%%%%%%%%%%%%%%%%%%%%%%%%%%%%%%%%%%%%%%%%%%
\section{Analytical expression of Green's function for free electron system in the RSFD approach\label{sec:appB}}
A 1-dimensional system is firstly considered for simplicity. In the RSFD approach, the kinetic-energy operator $\hat{K}=-\frac{1}{2}\nabla^2$ is represented by the matrix $\hat{K}^{(\mathcal{N}_f)}$, and the kinetic-energy term in the Kohn--Sham equation is written as
\begin{eqnarray}
\label{eqn:lsB-01}
\hat{K}^{(\mathcal{N}_f)} \psi(z_\ell) &\equiv& -\frac{1}{2h^2}\Bigl[ C_{-\mathcal{N}_f}\psi(z_{\ell-\mathcal{N}_f}) + C_{-\mathcal{N}_f+1}\psi(z_{\ell-\mathcal{N}_f+1}) + \cdots \Bigr. \nonumber \\ 
&&\hspace{1.5cm} \Bigl. \cdots + C_{-1}\psi(z_{\ell-1}) + C_{0}\psi(z_{\ell}) + C_{1}\psi(z_{\ell+1}) + \cdots \Bigr. \nonumber \\ 
&&\hspace{3cm} \Bigl. \cdots + C_{\mathcal{N}_f-1}\psi(z_{\ell+\mathcal{N}_f-1}) + C_{\mathcal{N}_f}\psi(z_{\ell+\mathcal{N}_f}) \Bigr],
\end{eqnarray}
where $\mathcal{N}_f$ is the order of the finite-difference approximation, $h$ is a grid spacing and the weight coefficients $C_{i}~(i=\ell-\mathcal{N}_f,\ell-\mathcal{N}_f+1,\cdots,\ell+\mathcal{N}_f)$ are determined using the Taylor expansion~\cite{fd}.

In the $\mathcal{N}_f$th order finite-difference approximation, the Green's function matrix $\hat{G}^{(\mathcal{N}_f)}$ is determined as satisfying 
\begin{eqnarray}
\label{eqn:lsB-02}
\left( Z- \hat{K}^{(\mathcal{N}_f)} \right) \hat{G}^{(\mathcal{N}_f)}(Z)&=& \hat{I},
\end{eqnarray}
where $Z$ is a complex number and $\hat{I}$ is the unit matrix. The $\ell$~th row--$\ell^\prime$ th column element of the Green's function matrix $G^{(\mathcal{N}_f)}(z_k,z_l;Z)$ is described in a spectral representation as 
\begin{eqnarray}
\label{eqn:lsB-03}
G^{(\mathcal{N}_f)}(z_k,z_l;Z)&=& \int^{\pi/h}_{-\pi/h}\frac{\phi_p(z_k)\phi^\ast_p(z_l)}{Z-E^{(\mathcal{N}_f)}_p}dp ,
\end{eqnarray}
where $E^{(\mathcal{N}_f)}_p$ and $\phi_p(z_{\ell})$ are the eigenvalue and eigenvector of $\hat{K}^{(\mathcal{N}_f)}$, respectively, obtained by solving the eigenvalue equation $\hat{K}^{(\mathcal{N}_f)}\phi_p(z_{\ell})=E^{(\mathcal{N}_f)}_p\phi_p(z_{\ell})$, and are given by
\begin{eqnarray}
\label{eqn:lsB-04}
\left\{
\begin{array}{rcl}
E^{(\mathcal{N}_f)}_p &=& \displaystyle{ -\frac{1}{2h^2}\left( C_0 + 2\sum^{\mathcal{N}_f}_{m=1} C_m\cos{mph} \right)} \\
\phi_p(z_\ell) &=& \displaystyle{ \sqrt{\frac{h}{2\pi}} \e^{ipz_\ell} }
\end{array}
\right. ,
\end{eqnarray}
where $-\frac{\pi}{h}<p\leq\frac{\pi}{h}$, and $\phi_p(z_\ell)$ is normalized, i.e.,
\begin{eqnarray}
\label{eqn:lsB-04-2}
\sum^{\infty}_{\ell=-\infty} \phi_p^\ast(z_\ell)\phi_{p^\prime}(z_\ell) &=& \delta(p-p^\prime).
\end{eqnarray}
Substituting Eq.~(\ref{eqn:lsB-04}) into Eq.~(\ref{eqn:lsB-03}) and changing the integration variable from $p$ to $\theta = ph$ and subsequently from $\theta$ to $\omega = \e^{i\theta}$, we obtain
\begin{eqnarray}
\label{eqn:lsB-05}
G^{(\mathcal{N}_f)}(z_k,z_l;Z) &=& \frac{h^2}{2\pi}\int^{\pi}_{-\pi}\frac{\e^{i\theta(k-l)}}{\displaystyle{ h^2Z+\frac{1}{2}C_0+\sum^{\mathcal{N}_f}_{m=1}C_m\cos{m\theta}}}d\theta \nonumber \\
&=& \frac{h^2}{2\pi i}\oint\frac{\omega^{|k-l|-1}}{\displaystyle{ h^2Z+\frac{1}{2}C_0+\frac{1}{2}\sum^{\mathcal{N}_f}_{m=1}C_m(\omega^m+\omega^{-m})}}d\omega.
\end{eqnarray}
The integration can be carried out along the unit circle in the complex $\omega$ plane based on the residue theorem. In the following, we introduce a sensible manner of picking up the poles inside the unit circle that contribute to the integration. These poles $\omega$'s are the $\mathcal{N}_f$ solutions of the equation
\begin{eqnarray}
\label{eqn:lsB-06}
h^2Z+\frac{1}{2}C_0+\frac{1}{2}\sum^{\mathcal{N}_f}_{m=1}C_m(\omega^m+\omega^{-m}) &=& 0.
\end{eqnarray}
We now define a new variable $s$ as
\begin{eqnarray}
\label{eqn:lsB-07}
s &=& \frac{1}{2}(\omega+\omega^{-1}) = \cos{\theta},
\end{eqnarray}
and rewrite Eq.~(\ref{eqn:lsB-06}) as
\begin{eqnarray}
\label{eqn:lsB-08}
h^2Z+\frac{1}{2}C_0+\sum^{\mathcal{N}_f}_{m=1}C_m\cos{m\theta} &=& 0,
\end{eqnarray}
which is the $\mathcal{N}_f$th order algebraic equation with respect to $s$ and its solutions are denoted by $s_n~(n=1,2,\cdots,\mathcal{N}_f)$. Resultantly, for each $s_n$, the poles $\omega_n$'s are given by the solutions of the quadratic equation $\omega^2_n-2s_n\omega_n+1=0$ as
\begin{eqnarray}
\label{eqn:lsB-09}
\omega_n^{(\pm)} &=& s_n \pm \sqrt{s_n^2-1}.
\end{eqnarray}
When the imaginary part of $Z$ is non-zero, either of $\omega^{(+)}_n$ or $\omega^{(-)}_n$ is inside the unit circle, while the other is outside it. Hereafter, the inside pole is defined as $\omega_n$, and the other is obtained by $\omega_n^{-1}$. Using the residue theorem, the integration of Eq.~(\ref{eqn:lsB-05}) is carried out to yield
\begin{eqnarray}
\label{eqn:lsB-10}
G^{(\mathcal{N}_f)}(z_k,z_l;Z) &=& \frac{2h^2}{C_{\mathcal{N}_f}}\sum^{\mathcal{N}_f}_{n=1}\frac{\omega_n^{|k-l|+\mathcal{N}_f-1}}{\displaystyle{ (\omega_n-\omega_n^{-1})\prod^{\mathcal{N}_f}_{m=1 \atop m\neq n}(\omega_n-\omega_m)(\omega_n-\omega_m^{-1})}} \nonumber \\
&=& \frac{h^2}{2^{\mathcal{N}_f-2}C_{\mathcal{N}_f}}\sum^{\mathcal{N}_f}_{n=1}\frac{\omega_n^{|k-l|}}{\displaystyle{ (\omega_n-\omega_m^{-1})\prod^{\mathcal{N}_f}_{m=1 \atop m\neq n}(s_n-s_m)}}.
\end{eqnarray}
Finally, introducing $\mathcal{K}_n$ by $s_n=\cos{\mathcal{K}_nh}$, hence, $\omega_n=\e^{i\mathcal{K}_nh}~({\rm Im}\{ \mathcal{K}_n \}>0)$, we obtain
\begin{eqnarray}
\label{eqn:lsB-11}
G^{(\mathcal{N}_f)}(z_k,z_l;Z) &=& \frac{h^2}{i2^{\mathcal{N}_f-1}C_{\mathcal{N}_f}}\sum^{\mathcal{N}_f}_{n=1}\frac{\e^{i\mathcal{K}_n|z_k-z_l|}}{\displaystyle{ \sin{\mathcal{K}_nh} \prod^{\mathcal{N}_f}_{m=1 \atop m\neq n}(\cos{\mathcal{K}_nh}-\cos{\mathcal{K}_mh})}}.
\end{eqnarray}
The retarded Green's function is given by
\begin{eqnarray}
\label{eqn:lsB-12}
G^{(\mathcal{N}_f)r}(z_k,z_l;E) &=& \lim_{\varepsilon \rightarrow 0^+}G^{(\mathcal{N}_f)}(z_k,z_l;E+i\varepsilon).
\end{eqnarray}

It is noted that $\mathcal{K}_n$ is a multivalued function since it is defined as $\mathcal{K}_n=\frac{1}{h}\cos^{-1}{s_n}$. The branch of $\cos^{-1}$ should be chosen to satisfy the requirement of ${\rm Im}\{ \mathcal{K}_n \}>0$, which guarantees that $\omega_n$ exists inside the unit circle in the complex $\omega$ plane. Thus, the following relationship is established:
\begin{eqnarray}
\label{eqn:lsB-13}
\left\{
\begin{array}{llcc}
\mbox{If} & {\rm Im}\{s_n\}<0, & \mbox{then} & \displaystyle{ 0<{\rm Re}\{\mathcal{K}_n\}<\frac{\pi}{h}} \\
\mbox{If} & {\rm Im}\{s_n\}>0, & \mbox{then} & \displaystyle{ -\frac{\pi}{h}<{\rm Re}\{\mathcal{K}_n\}<0}
\end{array}
\right. .
\end{eqnarray}

\noindent{\bf Proof:}
Consider complex numbers $s \equiv \xi+i\eta$ and $\mathcal{K} \equiv k+i\kappa$ with the relationship of $s=\cos{\mathcal{K}h}$ in the interval of $|k|<\frac{\pi}{h}$. It is readily shown that when $\eta<0$ and $\kappa>0$, there exists the one-to-one correspondence between $\{ s \}$ and $\{ \mathcal{K} \}$. In this case, $\mathcal{K}$ is uniquely so defined as to satisfy ${\rm Im}\{\mathcal{K}\}=\kappa>0$ and the relationship
\begin{eqnarray}
\label{eqn:lsB-14}
\left\{
\begin{array}{rcl}
k &=& \left\{
\begin{array}{lcl}
\displaystyle{\frac{1}{h}\cos^{-1}{  \gamma^{-} }} & \cdots & \xi \geq 0 \\ \\
\displaystyle{\frac{1}{h}\cos^{-1}{(-\gamma^{-})}} & \cdots & \xi <  0
\end{array}
\right. \\ \\
\kappa &=& \displaystyle{\frac{1}{h}\cosh^{-1}{\gamma^{+}}}
\end{array}
\right. ,
\end{eqnarray}
where $\gamma^{\pm}$ is defined as 
\begin{eqnarray}
\label{eqn:lsB-15}
\gamma^{\pm} &=& \sqrt{\frac{\xi^2+\eta^2+1\pm\sqrt{(\xi^2+\eta^2-1)^2+4\eta^2}}{2}},
\end{eqnarray}
and $\cos^{-1} (\cosh^{-1})$ in Eq.~(\ref{eqn:lsB-14}) is the principal value of the inverse trigonometric (hyperbolic) cosine function.
Thus, $k$ varies as
\begin{eqnarray}
\label{eqn:lsB-16}
\left\{
\begin{array}{lcl}
\displaystyle{             0<k\leq\frac{\pi}{2h}} & \cdots & \xi \geq 0 \\ \\
\displaystyle{\frac{\pi}{2h}<k<\frac{\pi}{ h}} & \cdots & \xi <  0
\end{array}
\right. .
\end{eqnarray}
In the derivation of Eqs.~(\ref{eqn:lsB-14})--(\ref{eqn:lsB-16}), we used well-known formulas,
\begin{eqnarray}
\label{eqn:lsB-17}
\left\{
\begin{array}{l}
\cos{(k+i\kappa)h} = \cos{kh}\cosh{\kappa{h}}-i\sin{kh}\sinh{\kappa{h}} \\
\sin^2{kh}+\cos^2{kh} = 1 \\
-\sinh^2{\kappa{h}}+\cosh^2{\kappa{h}} = 1 
\end{array}
\right. .
\end{eqnarray}

On the other hand, in the case of $\eta>0$ and $\kappa>0$, then $k$ and $\kappa$ are determined in the same manner as
\begin{eqnarray}
\label{eqn:lsB-18}
\left\{
\begin{array}{rcl}
k &=& \left\{
\begin{array}{lcl}
\displaystyle{-\frac{1}{h}\cos^{-1}{  \gamma^{-} }} & \cdots & \xi \geq 0 \\ \\
\displaystyle{-\frac{1}{h}\cos^{-1}{(-\gamma^{-})}} & \cdots & \xi <  0
\end{array}
\right. \\ \\
\kappa &=& \displaystyle{\frac{1}{h}\cosh^{-1}{\gamma^{+}}}
\end{array}
\right. ,
\end{eqnarray}
respectively, and $k$ varies as
\begin{eqnarray}
\label{eqn:lsB-19}
\left\{
\begin{array}{lcl}
\displaystyle{-\frac{\pi}{2h}\leq k<              0} & \cdots & \xi \geq 0 \\ \\
\displaystyle{-\frac{\pi}{ h}<k<-\frac{\pi}{2h}} & \cdots & \xi <  0
\end{array}
\right. .
\end{eqnarray}
\hfill(Q.E.D)

It is straightforward to extend the above argument to the 3-dimensional case. We deal with a free electron system in which the discretized space is infinite in the $z$ direction and periodic in the $x$ and $y$ directions.
The Green's function in this case is described in a spectral representation by
\begin{eqnarray}
\label{eqn:lsB-20}
G^{(\mathcal{N}_f)}(\vecvar{r}_{\dsla,j},z_k,\vecvar{r}_{\dsla,j^\prime},z_l;Z) &=& \sum^{\frac{N_x-1}{2}}_{n_x=-\frac{N_x-1}{2}} \sum^{\frac{N_y-1}{2}}_{n_y=-\frac{N_y-1}{2}} \int^{\frac{\pi}{h_z}}_{-\frac{\pi}{h_z}}{\frac{\phi_{n_x,n_y,p}(\vecvar{r}_{\dsla,j},z_k)~\phi^\ast_{n_x,n_y,p}(\vecvar{r}_{\dsla,j^\prime},z_l)}{Z-E^{(\mathcal{N}_f)}_{n_x,n_y,p}}}dp, \nonumber \\
\end{eqnarray}
where $E^{(\mathcal{N}_f)}_{n_x,n_y,p}$ and $\phi_{n_x,n_y,p}(\vecvar{r}_{\dsla,j},z_{\ell})$ are the eigenvalue and eigenvector of the 3-dimensional kinetic-energy matrix, respectively, i.e.,
\begin{eqnarray}
\label{eqn:lsB-21}
\left\{
\begin{array}{l}
E^{(\mathcal{N}_f)}_{n_x,n_y,p} = E^{(\mathcal{N}_f)}_{n_x,n_y} + E^{(\mathcal{N}_f)}_{p} \\
\left\{
\begin{array}{rcl}
E^{(\mathcal{N}_f)}_{n_x,n_y} &=& \displaystyle{-\frac{1}{2h_x^2}\Biggl(C_0+2\sum^{\mathcal{N}_f}_{m=1}C_m\cos{mG_{n_x}h_x} \Biggr)} \\
                              & & \displaystyle{-\frac{1}{2h_y^2}\Biggl(C_0+2\sum^{\mathcal{N}_f}_{m=1}C_m\cos{mG_{n_y}h_y} \Biggr)} \\
E^{(\mathcal{N}_f)}_{p}       &=& \displaystyle{-\frac{1}{2h_z^2}\Biggl(C_0+2\sum^{\mathcal{N}_f}_{m=1}C_m\cos{mph_z} \Biggr)}
\end{array}
\right. \\
\phi_{n_x,n_y,p}(\vecvar{r}_{\dsla,j},z_{\ell}) = \displaystyle{\sqrt{\frac{h_z}{2\pi N_xN_y}}\exp\Bigl(i\vecvar{G}_{\dsla,n}\cdot\vecvar{r}_{\dsla,j}+ipz_\ell\Bigr)}
\end{array}
\right. .
\end{eqnarray}
Here, $\vecvar{r}_{\dsla,j}=(x_{j_x},y_{j_y})$ are the lateral coordinates with $j_{x(y)}=1,2,\cdots,N_{x(y)}$ [for convenience, $N_{x(y)}$ is chosen an odd integer], and the definition of $\vecvar{G}_{\dsla,n}$ is same as shown in Eq.~(\ref{eqn:ls11}).
Now, the Green's function represented by Eq.~(\ref{eqn:lsB-11}) reads as
\begin{eqnarray}
\label{eqn:lsB-22}
G^{(\mathcal{N}_f)}(\vecvar{r}_{\dsla,j},z_k,\vecvar{r}_{\dsla,j^\prime},z_l;Z) &=& \frac{h_z^2}{i2^{\mathcal{N}_f-1}N_xN_yC_{\mathcal{N}_f}} \nonumber \\
& \times & \sum^{\frac{N_x-1}{2}}_{n_x=-\frac{N_x-1}{2}} \sum^{\frac{N_y-1}{2}}_{n_y=-\frac{N_y-1}{2}} \sum^{\mathcal{N}_f}_{n=1}\frac{\exp\Bigl(i\vecvar{G}_{\dsla,n}\cdot(\vecvar{r}_{\dsla,j}-\vecvar{r}_{\dsla,j^\prime})+i\mathcal{K}_n|z_k-z_l|\Bigr)}{\displaystyle{ \sin{\mathcal{K}_nh_z} \prod^{\mathcal{N}_f}_{m=1 \atop m\neq n}(\cos{\mathcal{K}_nh_z}-\cos{\mathcal{K}_mh_z})}}, \nonumber \\
\end{eqnarray}
where $\mathcal{K}_n=\frac{1}{h_z}\cos^{-1}{s_n}$ under the requirement of ${\rm Im}\{ \mathcal{K}_n \}>0$, and $s_n$ is the solution of the $\mathcal{N}_f$th order algebraic equation with respect to $s~(=\cos{\theta})$
\begin{eqnarray}
\label{eqn:lsB-23}
h_z^2\Bigl(Z-E^{(\mathcal{N}_f)}_{n_x,n_y}\Bigr)+\frac{1}{2}C_0+\sum^{\mathcal{N}_f}_{m=1}C_m\cos{m\theta} &=& 0.
\end{eqnarray}

In the following, we present the {\it analytic} representation of the retarded Green's functions in $\mathcal{N}_f=1 \sim 4$ cases in the 1-dimensional free electron system; there exists no analytic one in a case of $\mathcal{N}_f \ge 5$, since the algebraic equation (\ref{eqn:lsB-08}) with $\mathcal{N}_f \ge 5$ can not be solvable analytically according to Galois theory.
Given the solutions of Eq.~(\ref{eqn:lsB-08}), $s_n \equiv \xi_n+i\eta_n~(n=1,2,\cdots,\mathcal{N}_f)$, $K_n \equiv k_n+i\kappa_n$ are determined from Eqs.~(\ref{eqn:lsB-13})--(\ref{eqn:lsB-19}), and finally we obtain the analytic representation of the Green's function Eq.~(\ref{eqn:lsB-11}).
Hereafter, we choose $Z=E+i\varepsilon$~($\varepsilon$: an infinitesimal positive number) in Eq.~(\ref{eqn:lsB-11}) so as to deal with the retarded Green's function.

\subsection{case of central finite difference~($\mathcal{N}_f=1$)}
Substituting $C_0=-2$ and $C_1=1$ into Eq.~(\ref{eqn:lsB-08}), we have the equation
\begin{eqnarray}
\label{eqn:lsB-24}
s-1+h^2(E+i\varepsilon) &=& 0,
\end{eqnarray}
and its solution
\begin{eqnarray}
\label{eqn:lsB-25}
s_1 &\equiv& \xi_1+i\eta_1 = 1-h^2E-ih^2\varepsilon.
\end{eqnarray}
Since Eq.~(\ref{eqn:lsB-25}) indicates that $\eta_1 \rightarrow 0^{-}$ (an infinitesimal {\it negative} number) in the limit of $\varepsilon \rightarrow 0^{+}$, $\mathcal{K}_1 \equiv k_1+i\kappa$ is determined from Eqs.~(\ref{eqn:lsB-14}) and (\ref{eqn:lsB-15}) such that
\begin{eqnarray}
\label{eqn:lsB-26}
\left\{
\begin{array}{lcllcl}
k_1 = \displaystyle{\frac{1}{h}\cos^{-1}\Bigl(1-h^2E\Bigr)}& , & \kappa_1 = 0 & \cdots & 0 \le E < \displaystyle{\frac{2}{h^2}} \\
k_1 = 0                                                    & , & \kappa_1 = \displaystyle{\frac{1}{h_{}}\cosh^{-1}\bigl( 1-h^2E\bigr)} & \cdots & E < 0 \\ 
k_1 = \displaystyle{\frac{\pi}{h}}                         & , & \kappa_1 = \displaystyle{\frac{1}{h}\cosh^{-1}\bigl(-1+h^2E\bigr)} & \cdots & \displaystyle{\frac{2}{h^2}}  \le  E
\end{array}
\right. .
\end{eqnarray}

\subsection{case of 5-point finite difference~($\mathcal{N}_f=2$)}
Substituting $C_0=-5/2$, $C_1=4/3$ and $C_2=-1/12$ into Eq.~(\ref{eqn:lsB-08}) yields the quadratic equation with respect to $s$,
\begin{eqnarray}
\label{eqn:lsB-27}
s^2-8s+7-6h^2(E+i\varepsilon) &=& 0.
\end{eqnarray}
This equation has two solutions $s_1$ and $s_2$ given by
\begin{eqnarray}
\label{eqn:lsB-28}
s_1 &\equiv& \xi_1+i\eta_1 = 4-\sqrt{9+6h^2(E+i\varepsilon)}, \\
\mbox{where} & & \left\{
\begin{array}{lclcl}
\xi_1 = 4-\sqrt{9+6h^2E} & , & \eta_1 \rightarrow 0^{-}       & \cdots& \displaystyle{-\frac{3}{2h^2}} \le E \\
\xi_1 = 4                & , & \eta_1 = -\sqrt{|9+6h^2E|} < 0 & \cdots& E < \displaystyle{-\frac{3}{2h^2}} 
\end{array}
\right. , \nonumber 
\end{eqnarray}
and
\begin{eqnarray}
\label{eqn:lsB-29}
s_2 &\equiv& \xi_2+i\eta_2 = 4+\sqrt{9+6h^2(E+i\varepsilon)}, \\
\mbox{where} & & \left\{
\begin{array}{lclcl}
\xi_2 = 4+\sqrt{9+6h^2E} & , & \eta_2 \rightarrow 0^{+}      & \cdots& \displaystyle{-\frac{3}{2h^2}} \le E \\
\xi_2 = 4                & , & \eta_2 = \sqrt{|9+6h^2E|} > 0 & \cdots& E < \displaystyle{-\frac{3}{2h^2}} 
\end{array}
\right. . \nonumber 
\end{eqnarray}
Subsequently, from Eqs.~(\ref{eqn:lsB-14}), (\ref{eqn:lsB-15}) and (\ref{eqn:lsB-18}), $\mathcal{K}_n \equiv k_n + i\kappa_n$ can be described in an analytic form.

\subsection{case of 7-point finite difference~($\mathcal{N}_f=3$)}
The substitution of $C_0=-49/18$, $C_1=3/2$, $C_2=-3/20$ and $C_3=1/90$ into Eq.~(\ref{eqn:lsB-08}) leads to the cubic equation with respect to $s$,
\begin{eqnarray}
\label{eqn:lsB-30}
s^3-\frac{27}{4}s^2+33s-\frac{109}{4}+\frac{45}{2}h^2(E+i\varepsilon) &=& 0,
\end{eqnarray}
whose solutions are determined according to Cardano's formula as
\begin{eqnarray}
&&
\label{eqn:lsB-31}
\begin{array}{rclclclcl}
s_1 &\equiv& \xi_1+i\eta_1, & \mbox{  where  } & \displaystyle{\xi_1 = \frac{9}{4}-\tau^{+}+\tau^{-}} & \mbox{  and  } & \eta_1 \rightarrow 0^{-},
\end{array} \\
&&
\label{eqn:lsB-32}
\begin{array}{rclclclcl}
s_2 &\equiv& \xi_2+i\eta_2, & \mbox{  where  } & \displaystyle{\xi_2 = \frac{\tau^{+}-\tau^{-}}{2}} & \mbox{  and  } & \displaystyle{\eta_2 = \frac{\tau^{+}+\tau^{-}}{2}}>0,
\end{array} \\
&&
\label{eqn:lsB-33}
\begin{array}{rclclclcl}
s_3 &\equiv& \xi_3+i\eta_3, & \mbox{  where  } & \displaystyle{\xi_3 = \frac{\tau^{+}-\tau^{-}}{2}} & \mbox{  and  } &\displaystyle{\eta_3 =-\frac{\tau^{+}+\tau^{-}}{2}}<0.
\end{array}
\end{eqnarray}
Here,
\begin{eqnarray}
\label{eqn:lsB-34}
\tau^{\pm} &=& \frac{1}{4}\sqrt[^{^{^3}}\!]{5 \Bigl\{ \sqrt{(144h^2E+155)^2 + 5 \times 19^3} \pm (144h^2E+155) \Bigr\}},
\end{eqnarray}
and $\mathcal{K}_n \equiv k_n + i\kappa_n$ can be analytically represented using Eqs.~(\ref{eqn:lsB-14}), (\ref{eqn:lsB-15}) and (\ref{eqn:lsB-18}).

\subsection{case of 9-point finite difference~($\mathcal{N}_f=4$)}
Now, Substituting $C_0=-205/72$, $C_1=8/5$, $C_2=-1/5$, $C_3=8/315$ and $C_4=-1/560$ into Eq.~(\ref{eqn:lsB-08}), we obtain the quartic equation with respect to $s$,
\begin{eqnarray}
\label{eqn:lsB-36}
s^4-\frac{64}{9}s^3+27s^2-\frac{320}{3}s+\frac{772}{9}-70h^2(E+i\varepsilon) &=& 0.
\end{eqnarray}
Ferrari's solutions to Eq.~(\ref{eqn:lsB-36}) are utilized.
After tedious but straightforward calculations, we have the following representations:
\begin{eqnarray}
\label{eqn:lsB-37}
s_1 &\equiv& \xi_1+i\eta_1, \\ && \mbox{\!\!\!where} \left\{
\begin{array}{lclcl}
\displaystyle{\xi_1 = \frac{16}{9}+\sigma-\sqrt{-\sigma^2-\alpha+\frac{\beta}{\sigma}}}& , & \eta_1 \rightarrow 0^{-} & \cdots & E_0 \leq E \\
\displaystyle{\xi_1 = \frac{16}{9}+\sigma} & , & \eta_1 = \displaystyle{-\sqrt{\sigma^2+\alpha-\frac{\beta}{\sigma}}}<0 & \cdots & E < E_0
\end{array}
\right. \!\!\!, \nonumber \\
\label{eqn:lsB-38}
s_2 &\equiv& \xi_2+i\eta_2, \\ && \mbox{\!\!\!where} \left\{
\begin{array}{lclcl}
\displaystyle{\xi_2 = \frac{16}{9}+\sigma+\sqrt{-\sigma^2-\alpha+\frac{\beta}{\sigma}}}& , & \eta_2 \rightarrow 0^{+} & \cdots & E_0 \leq E \\
\displaystyle{\xi_2 = \frac{16}{9}+\sigma} & , & \eta_2 = \displaystyle{\sqrt{\sigma^2+\alpha-\frac{\beta}{\sigma}}}>0 & \cdots & E < E_0
\end{array}
\right. \!\!\!, \nonumber \\
\label{eqn:lsB-39}
s_3 &\equiv& \xi_3+i\eta_3, \\ && \mbox{\!\!\!where} \left\{
\begin{array}{lclcl}
\displaystyle{\xi_3 = \frac{16}{9}-\sigma} & , & \eta_3 = \displaystyle{-\sqrt{\sigma^2+\alpha+\frac{\beta}{\sigma}}}<0 & \cdots & E_0^{\prime} \leq E \\
\displaystyle{\xi_3 = \frac{16}{9}-\sigma} & , & \eta_3 = \displaystyle{ \sqrt{\sigma^2+\alpha+\frac{\beta}{\sigma}}}>0 & \cdots & E < E_0^{\prime}
\end{array}
\right. \!\!\!, \nonumber \\
\label{eqn:lsB-40}
s_4 &\equiv& \xi_4+i\eta_4, \\ && \mbox{\!\!\!where} \left\{
\begin{array}{lclcl}
\displaystyle{\xi_4 = \frac{16}{9}-\sigma}& , & \eta_4 = \displaystyle{ \sqrt{\sigma^2+\alpha+\frac{\beta}{\sigma}}}>0 & \cdots & E_0^{\prime} \leq E \\
\displaystyle{\xi_4 = \frac{16}{9}-\sigma}& , & \eta_4 = \displaystyle{-\sqrt{\sigma^2+\alpha+\frac{\beta}{\sigma}}}<0 & \cdots & E < E_0^{\prime}
\end{array}
\right. \!\!\!. \nonumber
\end{eqnarray}
Here,
\begin{eqnarray}
\left\{
\begin{array}{l}
\begin{array}{rclcrclc}
\label{eqn:lsB-41}
\alpha &=& \displaystyle{\frac{7 \cdot 31}{2 \cdot 3^3}} &,& \beta &=& \displaystyle{\frac{2^3 \cdot 7 \cdot 181}{3^6_{}}} &,
\end{array} \\
\sigma = \left\{
\begin{array}{lclcl}
\displaystyle{\sqrt{\alpha \left( -\frac{1}{3}+\sqrt[^{^3}\!]{\rho + \sqrt{\nu^3 + \rho^2}} - \sqrt[^{^3}\!]{-\rho + \sqrt{\nu^3 + \rho^2}} \right)}}& \cdots & 0 \leq \nu \\
\displaystyle{\sqrt{\alpha \left( -\frac{1}{3}+\sqrt[^{^3}\!]{\rho + \sqrt{\nu^3 + \rho^2}} + \sqrt[^{^3}\!]{ \rho - \sqrt{\nu^3 + \rho^2}} \right)}}& \cdots & \nu < 0
\end{array}
\right. , \\
\begin{array}{rclcrclc}
\nu &=& \displaystyle{a_0+a_1h^2E} &,& \rho &=& \displaystyle{\frac{1}{2_{}}\nu + a_2} &, \\
E_0 &=& \displaystyle{\frac{\nu_0 - a_0}{a_1h^2}} &,& E_0^\prime &=& \displaystyle{\frac{\nu^\prime_0 - a_0}{a_1h^2_{}}} &, \\
\end{array} \\
\begin{array}{rclcrclcrcl}
a_0 &=& \displaystyle{\frac{3^2 \cdot 5 \cdot 19}{2^2 \cdot 31^2}} &,& a_1 &=& \displaystyle{\frac{2 \cdot 3^5 \cdot 5}{7 \cdot 31^2}} &,& a_2 &=& \displaystyle{\frac{3^2 \cdot 5 \cdot 3623}{2 \cdot 7 \cdot 31^3}}, 
\end{array}
\end{array}
\right.
\end{eqnarray}
and $\nu_0$ is the solution of $\nu^3 + \rho^2 =0$, which is evaluated as $\nu_0 \neareq -0.3563$, and $\nu_0^\prime = -\Bigl( (a_3+1/3)^3-2a_2 \Bigr)/3a_3 \neareq -0.4132$ with $a_3$ being $\sqrt[{^3}]{2\beta^2}/2\alpha$.
The usage of Eqs.~(\ref{eqn:lsB-11})--(\ref{eqn:lsB-19}) leads to the analytic representations of $\mathcal{K}_n \equiv k_n + i\kappa_n$ and the retarded Green's function.

The above treatment for analytically describing the retarded Green's function is readily extendable to the 3-dimensional case using Eqs.~(\ref{eqn:lsB-21})--(\ref{eqn:lsB-23}).

%%%%%%%%%%%%%%%%%%%%%%%%%%%%%%%%%%%%%%%%%%%%%%%%%%%%%%%%%%%%%%%%%%%%%%%%%%
%!!!!!!!!!!!!!!!!!!!!!!!!!!!!!!!!!!!!!!!!!!!!!!!!!!!!!!!!!!!!!!!!!!!!
\newpage %Just because of unusual number of tables stacked at end

\begin{thebibliography}{99}
\bibitem{comm1}
An alternative derivation of Eq.~(\ref{eqn:ls22-3}) is given in Ref.~17, in which the retarded self-energy matrix of the jellium electrode, $\Sigma^{r}(z_k)$, is analytically calculated. See Eq.~(\ref{egami:eqn9-122}) for the relationship between $X_k$ and $\Sigma^r(z_k)$, and note also that $B(z_k)=-\frac{1}{2h_z^2}I$ in the central finite-difference ($\mathcal{N}_f=1$) case.%Ref.~17=\cite{iobm}
\bibitem{comment}
In each model, the leakage current was calculated by integrating the tunneling current over the energy ranges between VBM and CBM of the substrate.
\end{thebibliography}

%%%%%%%%%%%%%%%%%%%%%%%%%%%%%%%%%%%%%%%%%%%%%%%%%%%%%%%%%%%%%%%%%%%%%%%%%%
%!!!!!!!!!!!!!!!!!!!!!!!!!!!!!!!!!!!!!!!!!!!!!!!!!!!!!!!!!!!!!!!!!!!!
\newpage

%%%%%%%%%%%%%%%%%%%%%%%%%%%%%%%%%%%%%%%%%%%%%%%%%%%%%%%%%%%%%%%%%%%%%%%%%%%%%%%
%!!!!!!!!!!!!!!!!!!!!!!!!!!!!!!!!!!!!!!!!!!!!!!!!!!!!!!!!!!!!!!!!!!!!
\end{document}